\documentclass[10pt,a4paper]{article}
\usepackage{amsmath}
\usepackage{amsfonts}
\usepackage{amssymb}
\usepackage[dvips]{graphicx}
\usepackage{bm}

\newcommand{\be}{\begin{equation}}

\newcommand{\ee}{\end{equation}}

\newcommand{\ba}{\begin{array}}

\newcommand{\ea}{\end{array}}

\begin{document}

%\begin{flushright}
%CQUeST-2010-0371
%\end{flushright}

\title{\textbf{Dynamics of false vacuum bubbles\\ in Brans-Dicke theory}}
\author{\textsc{Bum-Hoon Lee}\footnote{bhl@sogang.ac.kr}\;, \textsc{Wonwoo Lee}\footnote{warrior@sogang.ac.kr}\\
\textit{Department of Physics and BK21 Division,}\\
\textit{and Center for Quantum Spacetime,}\\
\textit{Sogang University, Seoul 121-742, Republic of Korea}\\
and\\
\textsc{Dong-han Yeom}\footnote{innocent@muon.kaist.ac.kr} \\
\textit{Department of Physics, KAIST, Daejeon 305-701, Republic of Korea}}
\maketitle

\begin{abstract}
We study the dynamics of false vacuum bubbles in the Brans-Dicke theory of gravity by using the thin shell or thin wall approximation.
We consider a false vacuum bubble that has a different value for the Brans-Dicke field between the inside false vacuum region and the outside true vacuum region.
Within a certain limit of field values, the difference of field values makes the effective tension of the shell negative.
This allows new expanding false vacuum bubbles to be seen by the outside observer, which are disallowed in Einstein gravity.
\end{abstract}

\newpage
\tableofcontents
\newpage

\newpage

\section{\label{sec:int}Introduction}

One of the most important intuitions into the nature of gravity would have to be Mach's principle. Mach argued that the inertial properties and the gravitational properties of an object are determined by the general mass distribution of the entire universe. Mach's speculation leads to the idea that the gravitation constant $G$ may change with time due to the time dependence of the density and the size of the universe. Dirac suggested another alternative, that $G$ can vary with time in his large number hypothesis: $G\varpropto 1/t$. In his hypothesis, he noticed that there are coincidences among very large dimensionless numbers. In Dirac's cosmology, the reason for these large values is simply the fact that the universe is old \cite{wein01}.

A more theoretically reliable and one of the most studied modified theories of gravity may be the Brans-Dicke theory \cite{brdi01}. Brans and Dicke introduced a scalar field $\Phi$ that is related to gravity by $G=1/\Phi$ and suggested the following Lagrangian:
\begin{eqnarray}\label{eq:BD}
\mathcal{L} = \frac{1}{16\pi} \left( \Phi R - \frac{\omega}{\Phi}g^{\alpha\beta} \nabla_{\alpha}\Phi\nabla_{\beta}\Phi \right),
\end{eqnarray}
where $R$ is the Ricci scalar, $\Phi$ is the Brans-Dicke field, and $\omega$ is the Brans-Dicke coupling constant.

Although the Brans-Dicke theory was inspired by Mach's principle, in fact, the theory has deep fundamental bases in fields from string theory to cosmology. For example, the Brans-Dicke theory with $\omega=-1$ limit can be obtained in a weak coupling limit of dilaton gravity, in which the dilaton field is a direct consequence of string theory \cite{Gasperini:2007zz}.
Also, the Brans-Dicke theory can be obtained in a weak field limit of the Randall-Sundrum model \cite{Randall:1999ee}, where $\omega$ is sufficiently large on the positive tension brane and $\omega \gtrsim -1.5$ on the negative tension brane \cite{Garriga:1999yh, Fujii:2003pa}.
Moreover, the Brans-Dicke field can violate the null energy condition.
Thus, it is useful in studying the exotic matter that allows wormhole geometry \cite{Agnese:1995kd}.
In cosmology, there has been some discussion to the effect that the Brans-Dicke field can be a candidate for dark matter or dark energy \cite{Setare:2006yj}.

In this paper, we study the dynamics of false vacuum bubbles in the Brans-Dicke theory using the thin shell or thin wall approximation \cite{israel01, Coleman:1980aw, Sato:1981bf, Blau:1986cw, Aguirre:2005xs, Alberghi:1999kd, cham01}. The dynamics of bubbles in the Brans-Dicke theory using the thin shell approximation have already been discussed by some authors \cite{gz0}. However, the previous authors considered cases in which the Brans-Dicke field becomes continuous around the shell. It is obviously true that the field value should be continuous; however, if the field varies on the shell and the shell is sufficiently thin, field values of the inside and the outside of the shell do not necessarily have to be the same \cite{Coleman:1980aw, Blau:1986cw}. In this paper, we allow discontinuous field values between the inside and the outside of the shell, while maintaining the condition of the thin shell approximation.

In Brans-Dicke theory, the strength of gravity can be tuned by the Brans-Dicke field $\Phi$. We will tune the gravity of the inside of the shell, while that of the outside will be Einstein gravity. One possible expectation is that the difference of the Brans-Dicke field can violate the null energy condition \cite{Hwang:2010aj}. A similar situation was studied by previous authors using a non-minimally coupled scalar field. In \cite{lll2006}, the authors obtained an expanding false vacuum bubble without the initial singularity after a nucleation \cite{lll2008}. In those works, the effective gravitation constant is still maintained as positive, even if the effective tension is negative. On the other hand, the effect of non-minimal coupling is interpreted as an additional positive tension on the shell of a true vacuum bubble. The additional tension due to the non-minimal coupling is changed dynamically in the case of massive true vacuum bubbles. In this case, there exists a bound type solution, which contracts and eventually collapses to a black hole \cite{lll20082}. The breathing false vacuum bubble has also been studied and the possible origins of the energy contents of thin shells has been discussed \cite{gs2008}.

In the original Brans-Dicke theory, the Brans-Dicke field does not have a potential term. Thus, it is difficult to obtain a stable de Sitter space \cite{hskim01}. In this paper, we introduce a potential of the Brans-Dicke field. This potential makes it possible to find a field configuration in which the inside of the shell is a stable de Sitter space while the outside is a stable Schwarzschild space. By tuning the Brans-Dicke field, we can make the gravitation of the inside of the bubble relatively weaker or stronger than that of the outside.
This can give a negative tension to the shell. Then, the negative tension allows new solutions that were disallowed in Einstein gravity \cite{Blau:1986cw, Aguirre:2005xs}. The new solutions include an expanding solution for the outside observer. Since the null energy condition is violated, the initial state of the bubble does not necessarily have to be unbuildable \cite{Farhi:1986ty, Farhi:1989yr, Freivogel:2005qh}.

This paper is organized as follows:
in Section~\ref{sec:eofm}, we derive the junction conditions in the Brans-Dicke theory of gravity by following the method used in \cite{cham01};
in Section~\ref{sec:dyn}, the dynamics of false vacuum bubbles are studied and their causal structures are classified;
and in Section~\ref{sec:dis}, we discuss the physical meaning of new solutions and their physical implications.

\section{\label{sec:eofm}The equation of motion of thin shells}

\subsection{\label{sec:fandj}Field equations and junction conditions in Brans-Dicke theory}

To describe a space-time with two qualitatively different domains and the transition region is relatively thinner than domains, one may describe such system by two distinct strategies. The intuitive approach is this: first calculate a continuous field combination between two domains, and second approximate the transition region as a physical object, e.g., a thin shell. However, the technically simple approach is this: consider two discontinuous domains and input a domain wall as an independent object. The first approach is physically correct, but it is equivalent with the second approach when the transition region is sufficiently thinner than each domains so that the transition region behaves as a physical object. In this limit, we can consider domain walls or thin shells by a Nambu-Goto type action and this method was already applied by previous authors \cite{cham01}.

In this paper, we consider a space-time that is separated by a thin shell into two distinct four-dimensional regions, ${\mathcal M}^{+}$ (outside) and ${\mathcal M}^{-}$ (inside), with boundaries, $\Sigma^{+}$ and $\Sigma^{-}$, respectively. To obtain a single glued space-time $\mathcal M = {\mathcal M}^{+} \cup {\mathcal M}^{-}$, we demand that the boundaries be identified as follows: $\Sigma^{+} = \Sigma^{-} = \Sigma$. This system can be obtained after the nucleation of a false vacuum bubble in the Brans-Dicke theory of gravity. Actually, the Brans-Dicke field and the scalar field are continuous from inside to outside through the wall or shell. In the shell, the fields vary continuously between the true and false vacuum states \cite{kllly}. After the nucleation of a vacuum bubble we may employ the thin shell approximation \cite{Coleman:1980aw}. In the approximation, the shell can be considered as a singular surface in the sense that a non-vanishing surface energy density exists. In this paper, we will consider the system employed the thin shell approximation after the nucleation of a false vacuum bubble.

Let us consider the following action of Brans-Dicke theory with a potential term and a matter field:
\begin{eqnarray}
S &=& \int_{\mathcal M} \sqrt{-g} d^4 x \left[ \frac{1}{16\pi} \left( \Phi R - \omega g^{\alpha\beta} \frac{\nabla_{\alpha}\Phi
\nabla_{\beta}\Phi}{\Phi} - V_{1}(\Phi) \right) - \frac{1}{2}{\nabla^\alpha}\phi
{\nabla_\alpha}\phi -V_{2}(\phi) \right] \nonumber \\
&+& \oint_{\Sigma} \sqrt{-h} d^3 x
\left[ \frac{\Phi (K-K_o)}{8\pi} - {\hat U}_1(\Phi) -{\hat U}_2(\phi) \right],\label{eq:bdm-action}
\end{eqnarray}
where $g\equiv \mathrm{det} g_{\mu\nu}$, $\Phi$ is the Brans-Dicke field, $R$ is the Ricci curvature scalar, $\omega$ is the dimensionless Brans-Dicke coupling constant, $K$ and $K_{o}$ are traces of the extrinsic curvatures of $\Sigma$ in the metric $g_{\mu\nu}$ and $\eta_{\mu\nu}$, respectively, $V_{1,2}$ are potentials that couple to the Brans-Dicke field and the matter field, and the second term on the right-hand side has the boundary term \cite{ygh} of Brans-Dicke theory and a Nambu-Goto type action on the shell. Here, we use coordinates $(x^{0}, x^{1}, x^{2}, \eta)$ where $\Sigma$ is the surface with the parameter $\eta = \bar{\eta}$ and $h_{\mu \nu}$ is the three-metric which is defined on $\Sigma$. ${\hat U}_1(\Phi)$ and ${\hat U}_2(\phi)$ are functions of $\Phi$ and $\phi$ on the wall, respectively. They are continuously connected with the field in each bulk space-time. Note that the cases of Einstein gravity with minimal coupling \cite{israel01, Coleman:1980aw, Sato:1981bf, Blau:1986cw, Aguirre:2005xs, Alberghi:1999kd, cham01} and non-minimal coupling \cite{lll2006, lll2008, lll20082} have been studied.

We vary the action to obtain metric junction conditions.
The variation of the first term in the Brans-Dicke action for the bulk side ${\mathcal M}_{\pm}$ gives
\begin{eqnarray}
\int_{\mathcal M} d^4 x \delta \left[\sqrt{-g}\Phi R \right]
&=&
\int_{\mathcal M}\sqrt{-g} d^4 x \left[ R \delta\Phi +  \Phi \left( R_{\mu\nu}-\frac{1}{2}g_{\mu\nu}R \right)\delta g^{\mu\nu} -(\nabla_{\mu}\nabla_{\nu}\Phi-g_{\mu\nu}\nabla^{\alpha}\nabla_{\alpha}\Phi) \delta g^{\mu\nu} \right]
\nonumber\\
&+& \oint_{\Sigma} \sqrt{-h} d^3 x \left[(\nabla_{\mu}\delta g_{\alpha\nu}-\nabla_{\alpha}\delta g_{\mu\nu})+(\delta g_{\mu\nu} \nabla_{\alpha}\Phi-\delta g_{\alpha\nu}\nabla_{\mu}\Phi)\right] \Phi n^{\alpha}h^{\mu\nu},\label{eq:vari1}
\end{eqnarray}
and the variation of the second term gives
\begin{eqnarray}
\int_{\mathcal M} d^4 x \delta \left[\sqrt{-g} \frac{\omega}{\Phi}\nabla^{\alpha}\Phi \nabla_{\alpha}\Phi \right] & =&
\int_{\mathcal M}\sqrt{-g} d^4 x \left[ \frac{\omega}{2\Phi} g_{\mu\nu}\nabla^{\alpha}\Phi \nabla_{\alpha}\Phi -\frac{\omega}{\Phi} \nabla_{\mu}\Phi \nabla_{\nu}\Phi \right]\delta g^{\mu\nu} \nonumber\\
&+&
\int_{\mathcal M}\sqrt{-g} d^4 x \left[ \frac{2\omega}{\Phi} \nabla^{\alpha}\nabla_{\alpha}\Phi -\frac{\omega}{\Phi^2} \nabla^{\alpha}\nabla_{\alpha}\Phi \right]\delta\Phi \nonumber\\
&-& 2\oint_{\Sigma} \sqrt{-h} d^3 x \left[\frac{1}{\Phi}\nabla_{\alpha}\Phi \right] \omega n^{\alpha} \delta\Phi,\label{eq:vari2}
\end{eqnarray}
where a unit normal vector $n^{\alpha}$ points in the direction of increasing $\eta$ if $\Sigma$ is time-like and $n^{\alpha}$ is future-directed if $\Sigma$ is space-like.
The variation of the boundary term in the Brans-Dicke theory gives
\begin{eqnarray}
\oint_{\Sigma} d^3x \delta [\sqrt{-h}\Phi K] & = & \oint_{\Sigma} \sqrt{-h} d^3 x \left[ K\delta\Phi + \Phi( \frac{1}{2}Kh^{\mu\nu}
\delta g_{\mu\nu} -K^{\mu\nu} \delta g_{\mu\nu} - h^{\mu\nu} n^{\alpha} \nabla_{\mu} \delta g_{\alpha\nu} \right. \nonumber \\
\;\;\;\; &+& \left.
\frac{h^{\mu\nu}}{2}n^{\alpha}\nabla_{\alpha}\delta g_{\mu\nu} + \frac{1}{2} K n^{\mu}n^{\nu}\delta g_{\mu\nu} )\right],
\end{eqnarray}
and the variation of the wall action gives
\begin{equation}
\oint_{\Sigma} d^3 x \delta [\sqrt{-h} (\hat{U_1}(\Phi) +\hat{U_2}(\phi))] = \oint_{\Sigma}
\sqrt{-h} d^3 x \left[ \frac{h^{\mu\nu}}{2}(\hat{U_1}+\hat{U_2})\delta g_{\mu\nu} + \left( \frac{d\hat{U_1}}{d\Phi}\delta\Phi + \frac{d\hat{U_2}}{d\phi}\delta\phi \right)\right].
\end{equation}
Here, the normal vector can be defined as follows:
\begin{eqnarray}\label{eq:normal}
n^{\alpha}n_{\alpha} \equiv \epsilon = \left\{
\begin{array}{ll}
-1 & \textrm{if $\Sigma$ is space-like}, \\
+1 & \textrm{if $\Sigma$ is time-like}.
\end{array} \right.
\end{eqnarray}
The relation with the metric is $g_{\mu\nu}=h_{\mu\nu}+\epsilon n_{\mu}n_{\nu}$.

Now we summarize all equations and junction conditions of the Brans-Dicke theory.

\begin{enumerate}

\item
The bulk Einstein equations are
\begin{equation}\label{eq:Einstein}
R_{\mu\nu} - \frac{1}{2}g_{\mu\nu}R = \frac{8\pi}{\Phi} \left( T^{\mathrm{BD}}_{\mu\nu} +  T^{\mathrm{M}}_{\mu\nu} \right),
\end{equation}
where $T^{\mathrm{BD}}_{\mu\nu}$ is the energy-momentum tensor of the Brans-Dicke field
\begin{equation}\label{eq:T_BD}
T^{\mathrm{BD}}_{\mu\nu}= \frac{1}{8\pi} \left[ \frac{\omega}{\Phi} \left( \nabla_{\mu}\Phi \nabla_{\nu}\Phi-\frac{1}{2}g_{\mu\nu} \nabla^{\alpha}\Phi\nabla_{\alpha}\Phi\right) +
\left(\nabla_{\mu}\nabla_{\nu}\Phi-g_{\mu\nu}\nabla^{\alpha}\nabla_{\alpha}\Phi\right) \right] - g_{\mu\nu} \frac{V_{1}(\Phi)}{16\pi},
\end{equation}
and $T^{\mathrm{M}}_{\mu\nu}$ is the matter energy-momentum tensor,
\begin{equation}\label{eq:T_M}
T^{\mathrm{M}}_{\mu\nu}=\nabla_{\mu}\phi \nabla_{\nu}\phi - \frac{1}{2}g_{\mu\nu}\nabla^{\alpha}\phi \nabla_{\alpha}\phi - g_{\mu\nu}V_{2}(\phi).
\end{equation}

\item
The field equation of the Brans-Dicke field on the bulk is
\begin{equation}\label{eq:BD_bulk}
\nabla^{\alpha}\nabla_{\alpha}\Phi = \frac{8\pi}{2\omega +3} T^{\mathrm{M}} + \frac{1}{2\omega+3} \left( \Phi \frac{dV_{1}(\Phi)}{d\Phi}-2V_{1}(\Phi) \right),
\end{equation}
where $T^{\mathrm{M}}=T^{\mathrm{M}\alpha}_{~~~~\alpha}$.

\item
The field equation of the scalar field on the bulk is
\begin{equation}\label{eq:M_bulk}
\frac{1}{\sqrt{-g}} \partial_{\mu} [\sqrt{-g} g^{\mu\nu}\partial_{\nu}\phi] = \frac{d V_{2}(\phi)}{d\phi}.
\end{equation}

\item
The boundary condition of the Brans-Dicke field at the thin shell is
\begin{equation}\label{eq:BD_boundary}
\lim_{\epsilon \rightarrow 0} \left[ \frac{2\omega}{\Phi} n^{\alpha}\nabla_{\alpha}\Phi - 2K \right]_{\eta=\bar{\eta}+\epsilon}
 - \lim_{\epsilon \rightarrow 0} \left[ \frac{2\omega}{\Phi} n^{\alpha}\nabla_{\alpha}\Phi - 2K \right]_{\eta=\bar{\eta}-\epsilon}= -16\pi \frac{d\hat{U}_{1}}{d\Phi}.
\end{equation}

\item
The boundary condition of the scalar field at the thin shell is
\begin{equation}\label{eq:M_boundary}
\lim_{\epsilon \rightarrow 0} \left[ n^{\alpha} \nabla_{\alpha}\phi \right]_{\eta=\bar{\eta}+\epsilon} - \lim_{\epsilon \rightarrow 0} \left[ n^{\alpha} \nabla_{\alpha}\phi \right]_{\eta=\bar{\eta}-\epsilon} = -\frac{d\hat{U}_2}{d\phi}.
\end{equation}

\item
The modified Lanczos equation, i.e., the junction condition of the shell, is given by
\begin{eqnarray}\label{eq:Lanczos}
\lim_{\epsilon \rightarrow 0} \left[ \Phi(K_{\mu\nu}-Kh_{\mu\nu})+ 2\Phi n^{\alpha} \nabla_{\alpha}\Phi h_{\mu\nu} \right]_{\eta=\bar{\eta}+\epsilon} &-& \lim_{\epsilon \rightarrow 0} \left[ \Phi(K_{\mu\nu}-Kh_{\mu\nu})+ 2\Phi n^{\alpha} \nabla_{\alpha}\Phi h_{\mu\nu} \right]_{\eta=\bar{\eta}-\epsilon} \nonumber \\
&=& 8\pi (\hat{U}_{1} +\hat{U}_{2}) h_{\mu\nu}.
\end{eqnarray}
%where
%\begin{equation}\label{eq:vari1}
%[K] \equiv \lim_{\epsilon \rightarrow 0} K^{+}(\eta = \bar\eta + \epsilon) - K^{-}(\eta = \bar\eta - \epsilon).
%\end{equation}

\end{enumerate}

Here we adopt the notations and sign conventions that were used in \cite{misner}. The sign arises because we have chosen the convention that
$n^{\alpha}$ points towards the region of increasing $\eta$. The $\bar{\eta}$ is the location of the hypersurface. The signs ($+$)
and ($-$) represent exterior and interior space-time, respectively.

After plugging Equation~(\ref{eq:BD_boundary}) into Equation~(\ref{eq:Lanczos}), the junction condition becomes
\begin{equation}\label{eq:jcbd}
\Phi_{+}K^{+}_{\mu\nu} - \Phi_{-}K^{-}_{\mu\nu} = -4\pi (\hat{U}_{1} + \hat{U}_{2}) h_{\mu\nu} -
\left(\Phi_{+} \frac{d\Phi_{+}}{d\eta} - \Phi_{-}\frac{d\Phi_{-}}{d\eta} \right)h_{\mu\nu}.
\end{equation}
This condition connects the difference of embedding of $\Sigma$ between two space-times through the energy-momentum tensor of $\Sigma$. We will give the equation of motion of the shell in the next section.

We take ${\hat U}_{1}$ as a constant $\sigma_{1}$ which may be a function of $\omega$ \cite{gz0};
${\hat U}_{2}$ is also a constant $\sigma_{2}$, which may be a function of $\omega$ and $\Phi$, and $\sigma_{1} + \sigma_{2} =\sigma$.
Actually, $\sigma$ is from the contribution of the shell as surface energy density in a bounce solution.
In the shell, the field varies continuously between the true and the false vacuum states. If the thickness of the shell is small compared to the radius of the shell and the other length scales (e.g., the horizon size of the inside de Sitter space $l$ or the mass size of the outside Schwarzschild space $M$), we can use the thin shell approximation \cite{Coleman:1980aw}. In the framework of junction conditions, we can consider the shell as a singular surface in the sense that a non-vanishing positive surface energy density or surface tension $\sigma$ exists. Thus, it becomes a kind of surface layer. In this paper, we take the energy-momentum tensor as the form
\begin{equation}\label{eq:emtensor}
\frac{1}{\Phi} \left(T^{\mathrm{BD}}_{\mu\nu} +  T^{\mathrm{M}}_{\mu\nu}\right) =  (\mathrm{regular \; terms}) + \frac{1}{\Phi_{\Sigma}} S_{\mu\nu} \delta(\eta-\bar\eta),
\end{equation}
where $\Phi_{\Sigma}$ is the value of $\Phi$ on the shell and $S_{\mu\nu}=-\sigma h_{\mu\nu} (x^{i}, \eta=\bar{\eta})$. The energy-momentum tensor of the hypersurface can be defined as the integral over the surface $\Sigma$ in the limit as the thickness $\epsilon$ goes to zero:
\begin{equation}\label{eq:epsilon}
\frac{1}{\Phi_{\Sigma}} S_{\mu\nu} = \lim_{\epsilon \rightarrow 0} \int^{\bar\eta +
\epsilon}_{\bar\eta - \epsilon} \frac{1}{\Phi} \left( T^{\mathrm{BD}}_{\mu\nu} +  T^{\mathrm{M}}_{\mu\nu} \right) d\eta.
\end{equation}
The constant of proportionally in the shell action becomes $\sigma (={\hat U}_{1} +{\hat U}_{2})$ because the internal structure of the shell is neglected in the thin shell limit and $S_{\eta\eta}=S_{\eta i}=0$ in the present work.
Note that the energy-momentum tensor around the shell is approximately $(\Delta \Phi/\epsilon)^{2}$, where $\Delta \Phi$ is the difference of the Brans-Dicke field between the inside and the outside. Then, as long as we can fine-tune the potential, without changing vacuum energy, we can freely choose the tension of the shell $\sigma$, which is approximately $\Delta \Phi^{2}/\epsilon$. However, in this paper, we will fix the size of $\Delta \Phi$, and, hence, the choice of the tension of the shell may have a certain limitation to hold the thin shell condition.

\subsection{\label{sec:eofmfvb}The equation of motion of false vacuum bubbles}

\begin{figure}
\begin{center}
\includegraphics[scale=0.75]{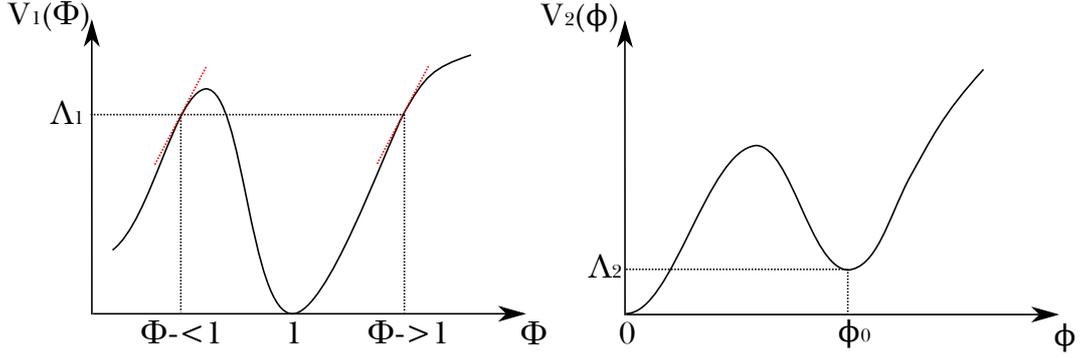}
\caption{\label{fig:potential} Required potential $V_{1}(\Phi)$ and $V_{2}(\phi)$.}
\end{center}
\end{figure}

We assume the following field configurations for the inside and the outside of the shell, where $R$ and $T$ are coordinates of the radius and the time with spherical symmetry, $\tau$ is the proper time measured by an observer at rest with respect to the shell and $r(\tau)$ is the proper circumferential radius of $\Sigma$:
\begin{enumerate}
\item The potential of the Brans-Dicke field $V_{1}(\Phi)$:
\begin{eqnarray}
V_{1}(\Phi) = \left\{ \begin{array}{ll}
0 & \Phi = 1,\\
\Lambda_{1} & \Phi = \Phi_{-}.
\end{array} \right.
\end{eqnarray}

\item The potential of the scalar field $V_{2}(\phi)$:
\begin{eqnarray}
V_{2}(\phi) = \left\{ \begin{array}{ll}
0 & \phi = 0,\\
\Lambda_{2} & \phi = \phi_{0}.
\end{array} \right.
\end{eqnarray}
We choose these shapes of potentials to make the outside a true vacuum with $\Phi=1$, while the inside is a false vacuum with $\Phi=\Phi_{-}$. See Figure~\ref{fig:potential}.

\item The Brans-Dicke field:
\begin{eqnarray}
\Phi(R,T) = \left\{ \begin{array}{ll}
1 & R > r(\tau),\\
\Phi_{-} & R < r(\tau).
\end{array} \right.
\end{eqnarray}

\item The scalar field:
\begin{eqnarray}
\phi(R,T) = \left\{ \begin{array}{ll}
0 & R > r(\tau),\\
\phi_{0} & R < r(\tau).
\end{array} \right.
\end{eqnarray}
Using these configurations, we induce a false vacuum region inside the shell. Due to the Brans-Dicke field, the strength of gravity is different between the inside and the outside of the shell.

\end{enumerate}

These constant field configurations can be shown to be a solution of field equations (Equations~(\ref{eq:BD_bulk}) and (\ref{eq:M_bulk})), if the following condition is satisfied:
\begin{equation}
V'_{1}(\Phi_{-})= \frac{32 \pi \Lambda_{\mathrm{eff}}}{\Phi_{-}},
\end{equation}
where
\begin{equation}
\Lambda_{\mathrm{eff}}=\frac{\Lambda_{1}}{16\pi}+\Lambda_{2} \equiv \frac{3}{8\pi l^{2}}.
\end{equation}

We can check that the following are solutions of Einstein equations in the inside and the outside of the shell:
\begin{equation}\label{eq:metric}
ds^{2} = -f_{\pm}(R) dT^{2} + \frac{dR^{2}}{f_{\pm}(R)} + R^{2} d\Omega^{2},
\end{equation}
where
\begin{equation}
f_{+}=1-\frac{2M}{R},
\end{equation}
and
\begin{equation}
f_{-}=1-\frac{R^2}{l^{2}\Phi_{-}}.
\end{equation}
Here, the induced metric on the shell can be written as
\begin{equation}
dS^2_{\Sigma} = - d\tau^2 + r^2(\tau) d\Omega^2.
\end{equation}
Since the induced metric must be the same on both sides of the shell, the following relation should be satisfied: $1=f_{\pm} {\dot T}^2 - f_{\pm}^{-1}{\dot R}^2$, where the dot is a derivation with respect to $\tau$.

With the ansatz, the second term on the right hand side of Equation~(\ref{eq:jcbd}) vanishes because $\frac{d\Phi_{+}}{d\eta}$ and
$\frac{d\Phi_{-}}{d\eta}$ vanish in the exterior and interior space-time of the shell, respectively in this paper.
The junction condition Equation~(\ref{eq:jcbd}) simplifies to
\begin{equation}\label{eq:jcbd2}
K^{+ i}_{\;\;\;j} - \Phi_-K^{-i}_{\;\;\;j}= - 4\pi \sigma \delta^{i}_{j}.
\end{equation}
Because of spherical symmetry, the extrinsic curvature has only two components, $K^{\theta}_{\theta} \equiv K^{\phi}_{\phi}$ and
$K^{\tau}_{\tau}$. The junction equation is related to $K^{\theta}_{\theta}$ and the covariant acceleration in the normal
direction is related to $K^{\tau}_{\tau}$.

The equation of motion for the shell, Equation~(\ref{eq:jcbd2}), becomes
\begin{equation}\label{eq:jc001}
\epsilon_{-} \Phi_{-}\sqrt{\dot{r}^2 + f_{-}} - \epsilon_{+} \sqrt{\dot{r}^2 + f_{+}} = 4\pi r \sigma,
\end{equation}
or, equivalently,
\begin{equation}\label{eq:jc002}
\epsilon_{-} \sqrt{\dot{r}^2 + f_{-}} - \epsilon_{+} \sqrt{\dot{r}^2 + f_{+}} = 4\pi r(\sigma + \bar\sigma),
\end{equation}
where
\begin{equation}
\bar\sigma = \epsilon_{-}\frac{1-\Phi_{-}}{4\pi r}\sqrt{\dot{r}^2 + f_{-}}
\end{equation}
represents the effect on the surface tension due to non-minimal coupling of the Brans-Dicke field.
$\epsilon_{\pm}$ are $+1$ if the normal vector on the shell is pointing towards increasing $r$ and $-1$ if towards decreasing $r$ \cite{Blau:1986cw}.

After squaring twice, we can get the equation as follows:
\begin{equation}\label{eq:emp01}
\frac{1}{2} {\dot r}^2 + V_{\mathrm{eff}}(r) = 0,
\end{equation}
where the effective potential is
\begin{equation}\label{eq:eff_pot}
V_{\mathrm{eff}}^{(1,2)}(r)= \frac{B\pm\sqrt{B^2 - AC}}{2A},
\end{equation}
with
\begin{eqnarray}
A&=&(\Phi_{-}^{2} -1)^2, \\
B&=&(\Phi_{-}^{2} +1)(\Phi^{2}_{-}f_{-} + f_{+} - 16\pi^{2}\sigma^{2}r^{2})-2\Phi^{2}_{-}(f_{+} +f_{-}), \\
C&=&(\Phi^{2}_{-}f_{-} + f_{+} - 16\pi^{2}\sigma^{2}r^{2})^{2} -4\Phi^{2}_{-} f_{+} f_{-}.
\end{eqnarray}
Since $B<0$, in the Einstein limit ($\Phi_{-}=1$), $V_{\mathrm{eff}}^{(1)}$ will converge to the Einstein limit
\begin{equation}
V_{\mathrm{eff}}^{\mathrm{E}}(r)= \frac{C}{4B},
\end{equation}
while $V_{\mathrm{eff}}^{(2)}$ diverges to $-\infty$.

\section{\label{sec:dyn}Dynamics of false vacuum bubbles in Brans-Dicke theory}

In this section, we study the dynamics of false vacuum bubbles. To define an effective potential, we consider four free parameters as follows: the size of the cosmological horizon of the inside de Sitter space $l$, the mass of the outside Schwarzschild space $M$, the tension $\sigma$, and the Brans-Dicke field of the inside $\Phi_{-}$. For convenience, we choose $l=2$. To test a positive $M$ case, we choose $M=0.5$ so that the size of the event horizon will be $2M=1$; also, we test $M=0$ case. The sign of $\bar{\sigma}$ depends on the sign of $1-\Phi_{-}$; hence, we observed two cases $\Phi_{-}>1$ and $\Phi_{-}<1$. $\sigma$ should be chosen to hold the thin shell approximation condition $\epsilon \sim (\Delta \Phi)^{2}/\sigma \ll l,M$.
\begin{table}[b]
\begin{center}
\begin{tabular}{c|c|c|c|c|c}
\hline
& \multicolumn{2}{c}{\;\;$\Phi_{-} = 1.001 > 1$\;\;} & \multicolumn{2}{|c|}{\;\;$\Phi_{-} = 0.999 < 1$\;\;} & \;\;$\Phi_{-} = 1.01 > 1$\;\; \\
\hline \hline
\;\;\;$\sigma$\;\;\; & \;\;\;\;$0.01$\;\;\;\; & \;\;\;\;$0.1$\;\;\;\; & \;\;\;\;$0.01$\;\;\;\; & \;\;\;\;$0.1$\;\;\;\; & $0.01$ \\
\hline
$M$ & $0.5$ & $0.5$ & $0.5$ & $0.5$ & $0$ \\
\hline
\end{tabular}
\caption{\label{table:condition}Initial conditions we used in this paper.}
\end{center}
\end{table}

\subsection{Conditions and results}

\begin{figure}
\begin{center}
\includegraphics[scale=0.5]{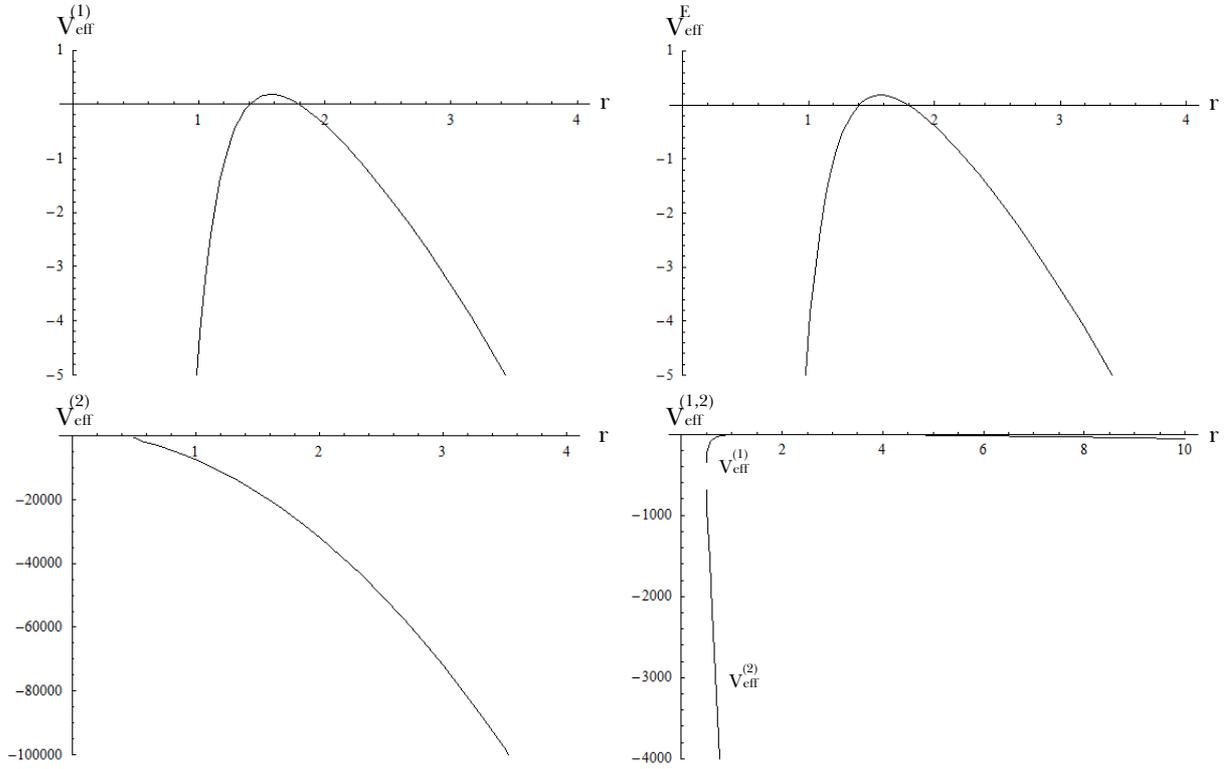}
\caption{\label{fig:more_V_001}The effective potentials $V_{\mathrm{eff}}^{(1)}$ (upper left), $V_{\mathrm{eff}}^{\mathrm{E}}$ (upper right), $V_{\mathrm{eff}}^{(2)}$ (lower left), and $V_{\mathrm{eff}}^{(1,2)}$ (lower right) for the $l=2$, $M=0.5$, $\sigma=0.01$, and $\Phi = 1.001$ case.}
\end{center}
\end{figure}
\begin{figure}
\begin{center}
\includegraphics[scale=0.5]{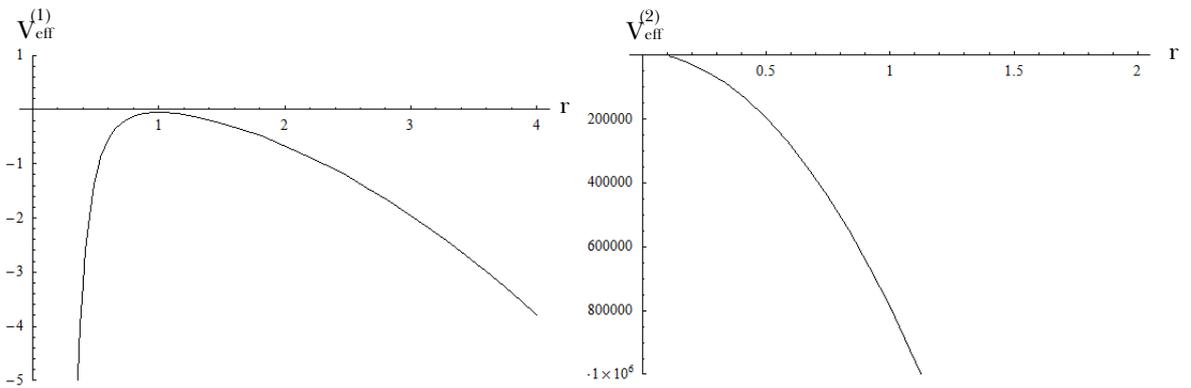}
\caption{\label{fig:more_V_01}The effective potentials $V_{\mathrm{eff}}^{(1)}$ (left) and $V_{\mathrm{eff}}^{(2)}$ (right) for the $l=2$, $M=0.5$, $\sigma=0.1$, and $\Delta \Phi = 0.001$ case.}
\end{center}
\end{figure}

First, we test the $\Phi_{-}>1$ case. We used $l=2$, $M=0.5$, $\sigma=0.01$, and $\Delta \Phi = 0.001$ (i.e., $\Phi_{-}=1.001$) so that
\begin{equation}
\epsilon \sim \frac{(\Delta \Phi)^{2}}{\sigma} \sim 10^{-4} \ll 1 \sim l,M.
\end{equation}
Hence, the thin shell approximation holds for these parameters.

We observe effective potentials $V_{\mathrm{eff}}^{(1)}$, $V_{\mathrm{eff}}^{(2)}$, and $V_{\mathrm{eff}}^{\mathrm{E}}$ (Figure~\ref{fig:more_V_001}). $V_{\mathrm{eff}}^{(1)}$ and $V_{\mathrm{eff}}^{\mathrm{E}}$ are similar around $V(r)\sim 0$.
If we observe more overall structures (lower right in Figure~\ref{fig:more_V_001}), we can see a difference between $V_{\mathrm{eff}}^{(1)}$ and $V_{\mathrm{eff}}^{\mathrm{E}}$. $V_{\mathrm{eff}}^{(1)}$ will end at a non-zero radius, since $B^{2}-AC$ becomes negative around there. Around that point, $V_{\mathrm{eff}}^{(2)}$ begins to decrease and decreases to negative infinity.

\begin{figure}
\begin{center}
\includegraphics[scale=0.5]{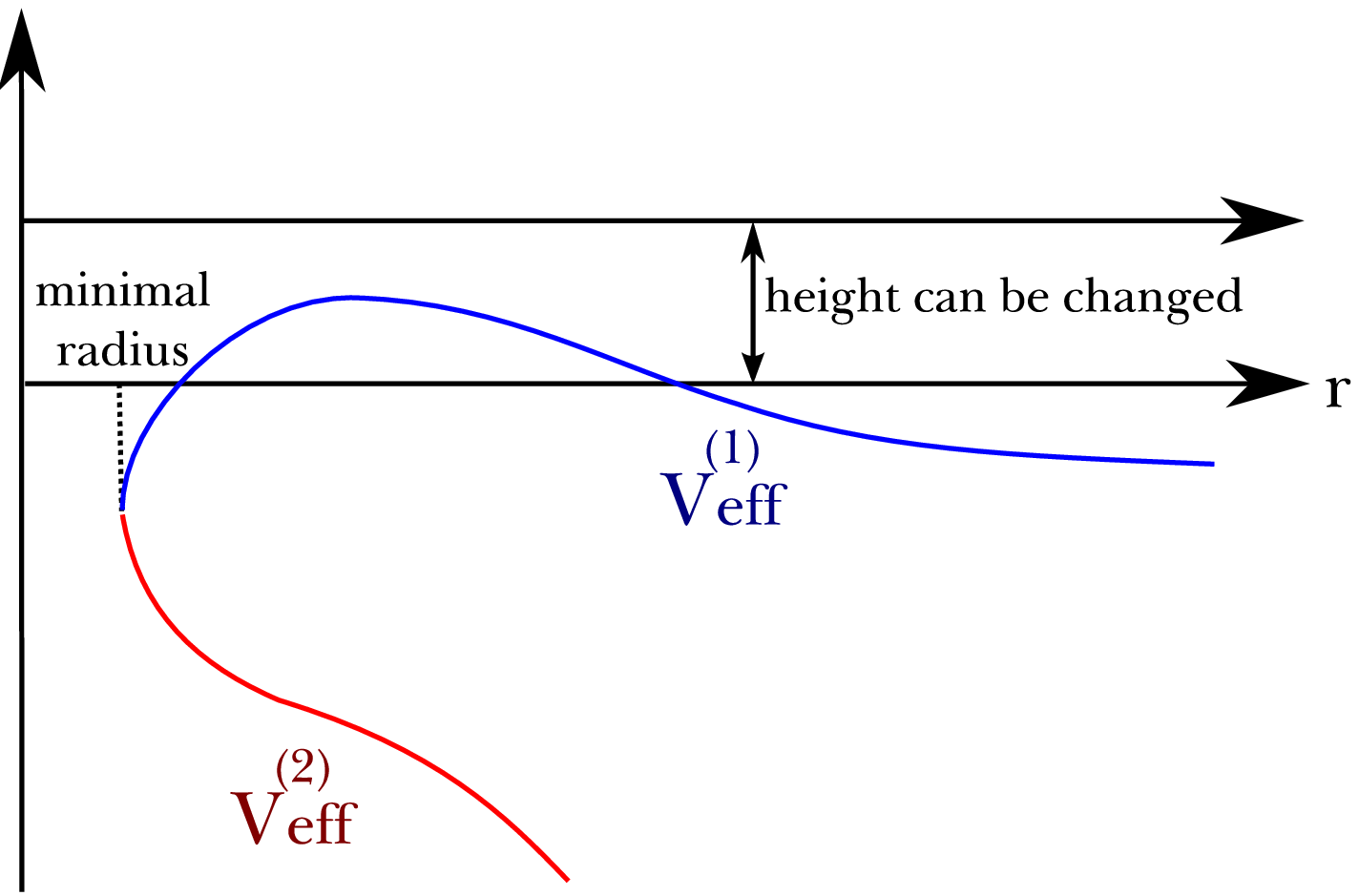}
\caption{\label{fig:eff_potential_more}The typical shapes of the effective potentials $V_{\mathrm{eff}}^{(1)}$ and $V_{\mathrm{eff}}^{(2)}$ for the $\Phi_{-}>1$. There is a minimal radius beyond which the thin shell approximation is not well-defined.}
\end{center}
\end{figure}

Also, we test $l=2$, $M=0.5$, $\sigma=0.1$, and $\Delta \Phi = 0.001$ (i.e., $\Phi_{-}=1.001$) so that
\begin{equation}
\epsilon \sim \frac{(\Delta \Phi)^{2}}{\sigma} \sim 10^{-5} \ll 1 \sim l,M.
\end{equation}
Here, we observed $V_{\mathrm{eff}}^{(1)}$ and $V_{\mathrm{eff}}^{(2)}$ again (Figure~\ref{fig:more_V_01}).
The effective potential $V_{\mathrm{eff}}^{(1)}$ is a convex function, while $V_{\mathrm{eff}}^{(2)}$ is a monotone function (Figure~\ref{fig:eff_potential_more}).

\begin{figure}
\begin{center}
\includegraphics[scale=0.5]{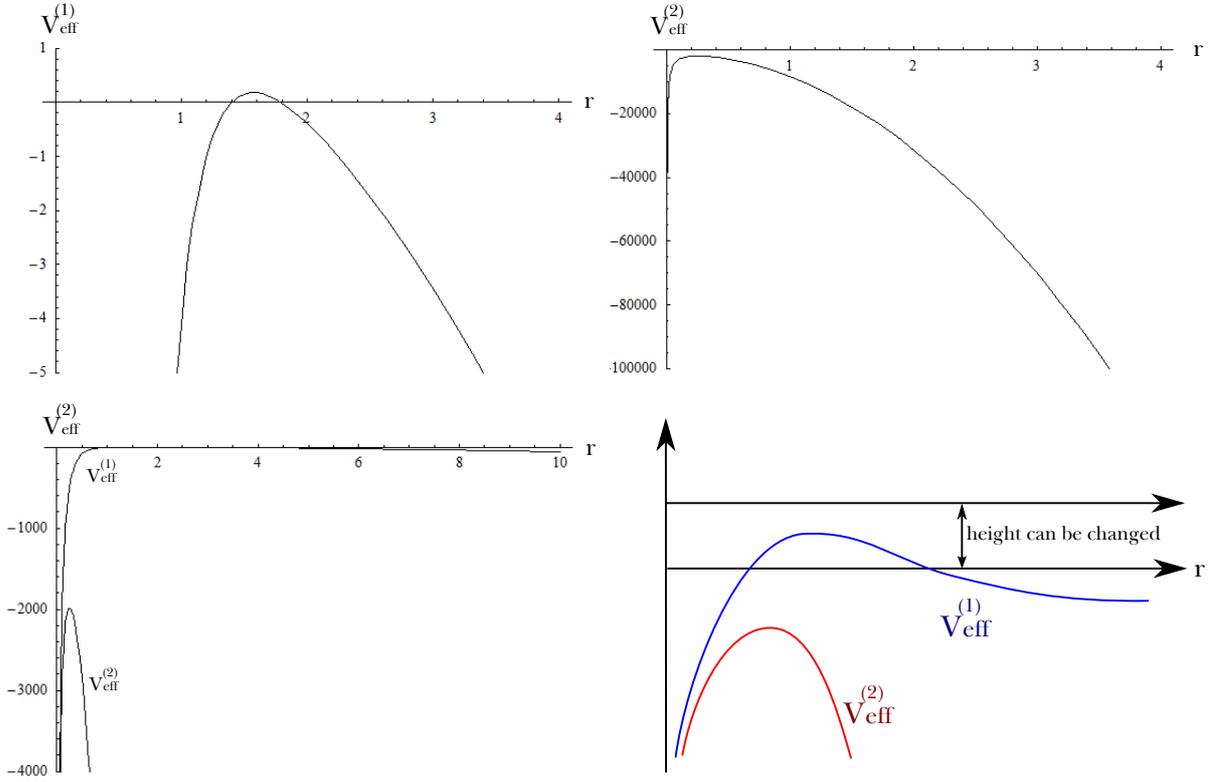}
\caption{\label{fig:less_V_001}The effective potentials $V_{\mathrm{eff}}^{(1)}$ (upper left), $V_{\mathrm{eff}}^{(2)}$ (upper right), $V_{\mathrm{eff}}^{(1,2)}$ (lower left), and their typical shapes (lower right) for the $l=2$, $M=0.5$, $\sigma=0.01$, and $\Phi = 0.999$ case.}
\end{center}
\end{figure}

\begin{figure}
\begin{center}
\includegraphics[scale=0.5]{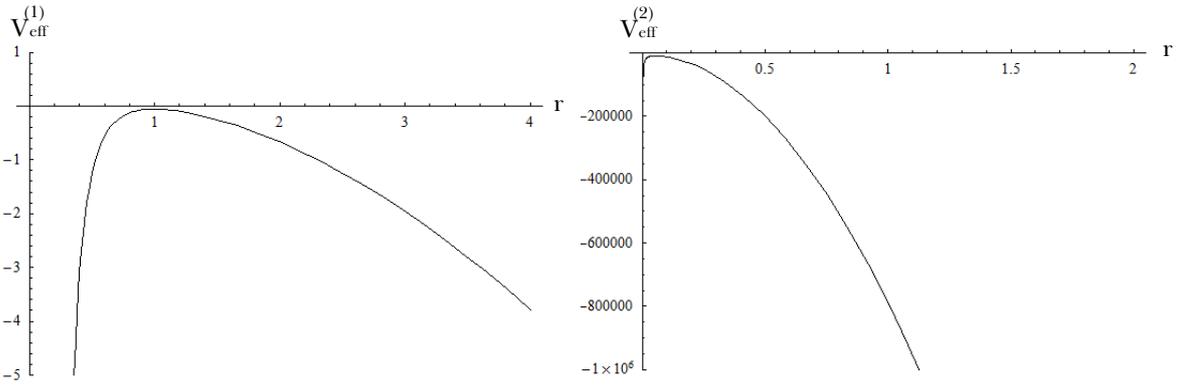}
\caption{\label{fig:less_V_01}The effective potentials $V_{\mathrm{eff}}^{(1)}$ (left) and $V_{\mathrm{eff}}^{(2)}$ (right) for the $l=2$, $M=0.5$, $\sigma=0.1$, and $\Delta \Phi = 0.999$ case.}
\end{center}
\end{figure}

Second, we test the $\Phi_{-}<1$ case. We used $l=2$, $M=0.5$, $\sigma=0.01$, and $\Delta \Phi = 0.001$ (i.e., $\Phi_{-}=0.999$).
We observe effective potentials $V_{\mathrm{eff}}^{(1)}$ and $V_{\mathrm{eff}}^{(2)}$, as shown in Figure~\ref{fig:less_V_001}.
Also, we test $l=2$, $M=0.5$, $\sigma=0.1$, and $\Delta \Phi = 0.001$ (i.e., $\Phi_{-}=0.999$), as shown in Figure~\ref{fig:less_V_01}.
The effective potentials $V_{\mathrm{eff}}^{(1,2)}$ are convex functions (lower right of Figure~\ref{fig:less_V_001}).

These effective potentials allow a collapsing solution or an expanding solution. Then, there are basically five possibilities: (a) from expanding to collapsing, (b) from collapsing to expanding, (c) from collapsing to collapsing, (d) from expanding to expanding, and (e) a static solution in an unstable equilibrium. (a) and (b) are symmetric solutions, whereas (c) and (d) are asymmetric solutions. For simplicity, we omit the unstable equilibrium case (e). Note that $V_{\mathrm{eff}}^{(2)}$ is only for asymmetric solutions.


\subsection{Determination of $\epsilon_{\pm}$ and the effective tensions}

Using the information for effective potentials, we know information for $\dot{r}^{2}$. Now we have to know further about the sign of $\epsilon_{\pm}$ to determine causal structures. Especially, the sign of $\epsilon_{-}$ is not entirely clear since it is combined with $\bar{\sigma}$. As in Einstein gravity, we define the effective extrinsic curvatures by the following form \cite{Blau:1986cw, Aguirre:2005xs, Freivogel:2005qh}:
\begin{eqnarray}
\beta^{(i)}_{-} = \frac{f_{-} - f_{+} + 16 \pi^{2} {\sigma^{(i)}_{\mathrm{eff}}}^{2} r^{2}}{8 \pi r \sigma^{(i)}_{\mathrm{eff}}} = \pm \sqrt{-2V^{(i)}_{\mathrm{eff}} + f_{-}}
\end{eqnarray}
and
\begin{eqnarray}
\beta^{(i)}_{+} = \frac{f_{-} - f_{+} - 16 \pi^{2} {\sigma^{(i)}_{\mathrm{eff}}}^{2} r^{2}}{8 \pi r \sigma^{(i)}_{\mathrm{eff}}} = \pm \sqrt{-2V^{(i)}_{\mathrm{eff}} + f_{+}},
\end{eqnarray}
where we define the \textit{effective tension}
\begin{eqnarray}
\sigma^{(i)}_{\mathrm{eff}}=\sigma+\bar{\sigma}^{(i)},
\end{eqnarray}
and $i=1,2$ denotes the index of the effective potential $V^{(1)}_{\mathrm{eff}}$ or $V^{(2)}_{\mathrm{eff}}$.
The effective extrinsic curvatures do not give direct results for the sign of each root because $\sigma_{\mathrm{eff}}$ already contains $\epsilon_{-}$. However, if we know $\epsilon_{-}$, then we can easily determine $\epsilon_{+}$ from the sign of $\beta_{+}$.

First, we summarize the sign of each root for the inside and the outside of the effective potentials when the effective tensions are always positive.
\begin{itemize}
\item $r\sim 0$ limit
\begin{itemize}
\item $\beta_{-}$ is always positive.
\item $\beta_{+}$ is always positive.
\end{itemize}
\item $r \rightarrow \infty$ limit
\begin{itemize}
\item $\beta_{-}$ can be positive or negative for a choice of parameters.
\item $\beta_{+}$ is always negative.
\end{itemize}
\end{itemize}

\begin{figure}
\begin{center}
\includegraphics[scale=0.45]{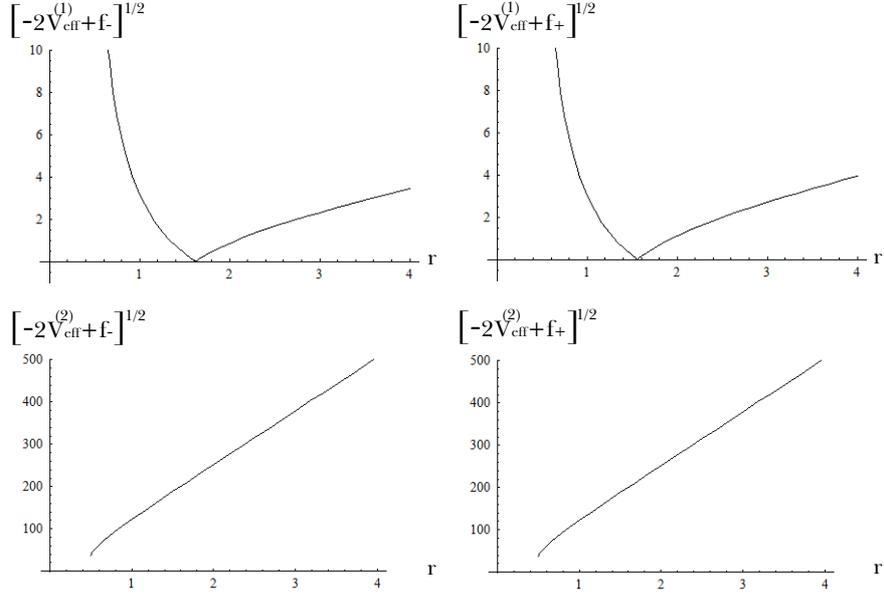}
\caption{\label{fig:more_sqrt_001}$\sqrt{\dot{r}^{2}+f_{-}}=\sqrt{-2V_{\mathrm{eff}}^{(1)}+f_{-}}$ (upper left), $\sqrt{-2V_{\mathrm{eff}}^{(1)}+f_{+}}$ (upper right), $\sqrt{-2V_{\mathrm{eff}}^{(2)}+f_{-}}$ (lower left), and $\sqrt{-2V_{\mathrm{eff}}^{(2)}+f_{+}}$ (lower right) for the $l=2$, $M=0.5$, $\sigma=0.01$, and $\Phi = 1.001$ case.}
\end{center}
\end{figure}

\begin{figure}
\begin{center}
\includegraphics[scale=0.45]{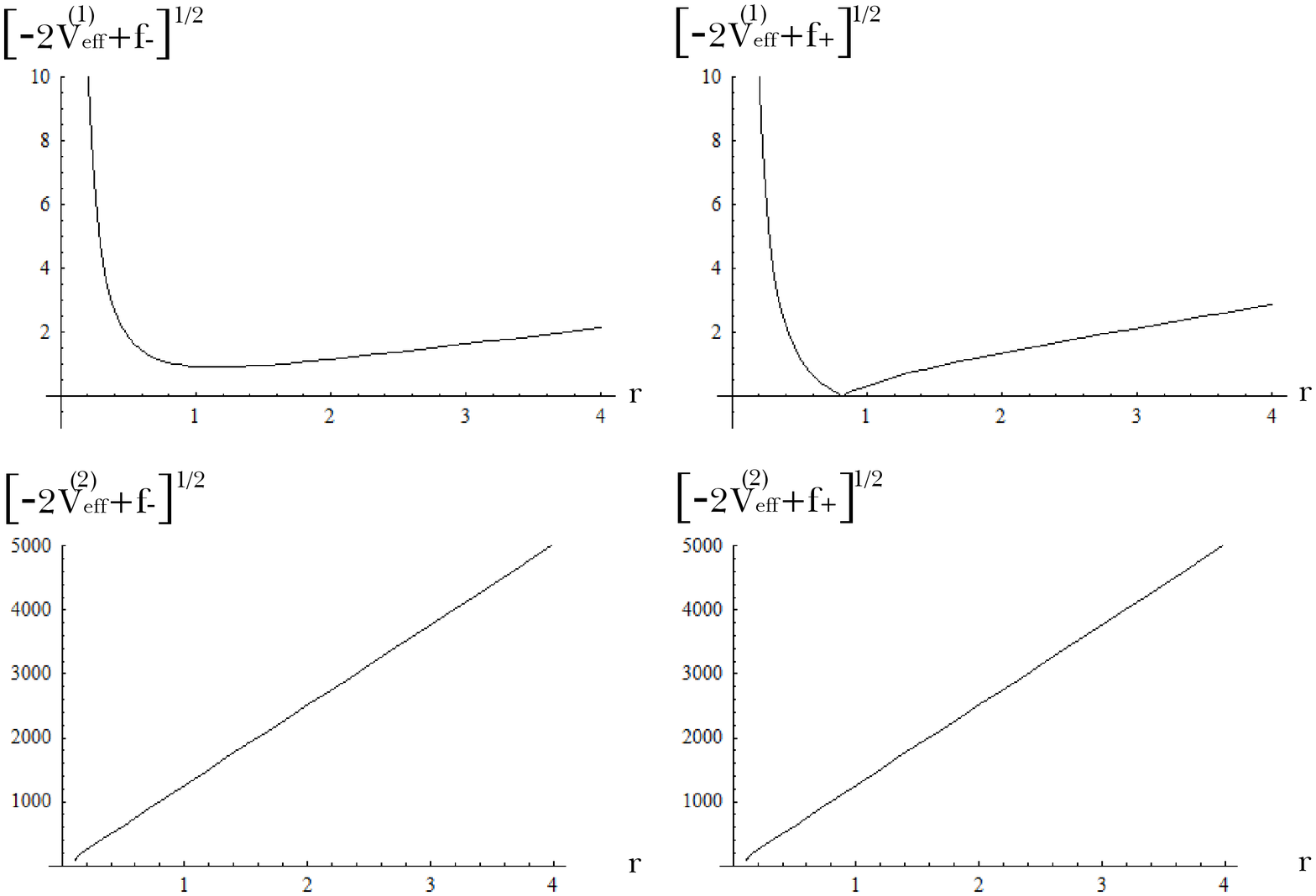}
\caption{\label{fig:more_sqrt_01}$\sqrt{-2V_{\mathrm{eff}}^{(1)}+f_{-}}$ (upper left), $\sqrt{-2V_{\mathrm{eff}}^{(1)}+f_{+}}$ (upper right), $\sqrt{-2V_{\mathrm{eff}}^{(2)}+f_{-}}$ (lower left), and $\sqrt{-2V_{\mathrm{eff}}^{(2)}+f_{+}}$ (lower right) for the $l=2$, $M=0.5$, $\sigma=0.1$, and $\Phi = 1.001$ case.}
\end{center}
\end{figure}

\begin{figure}
\begin{center}
\includegraphics[scale=0.45]{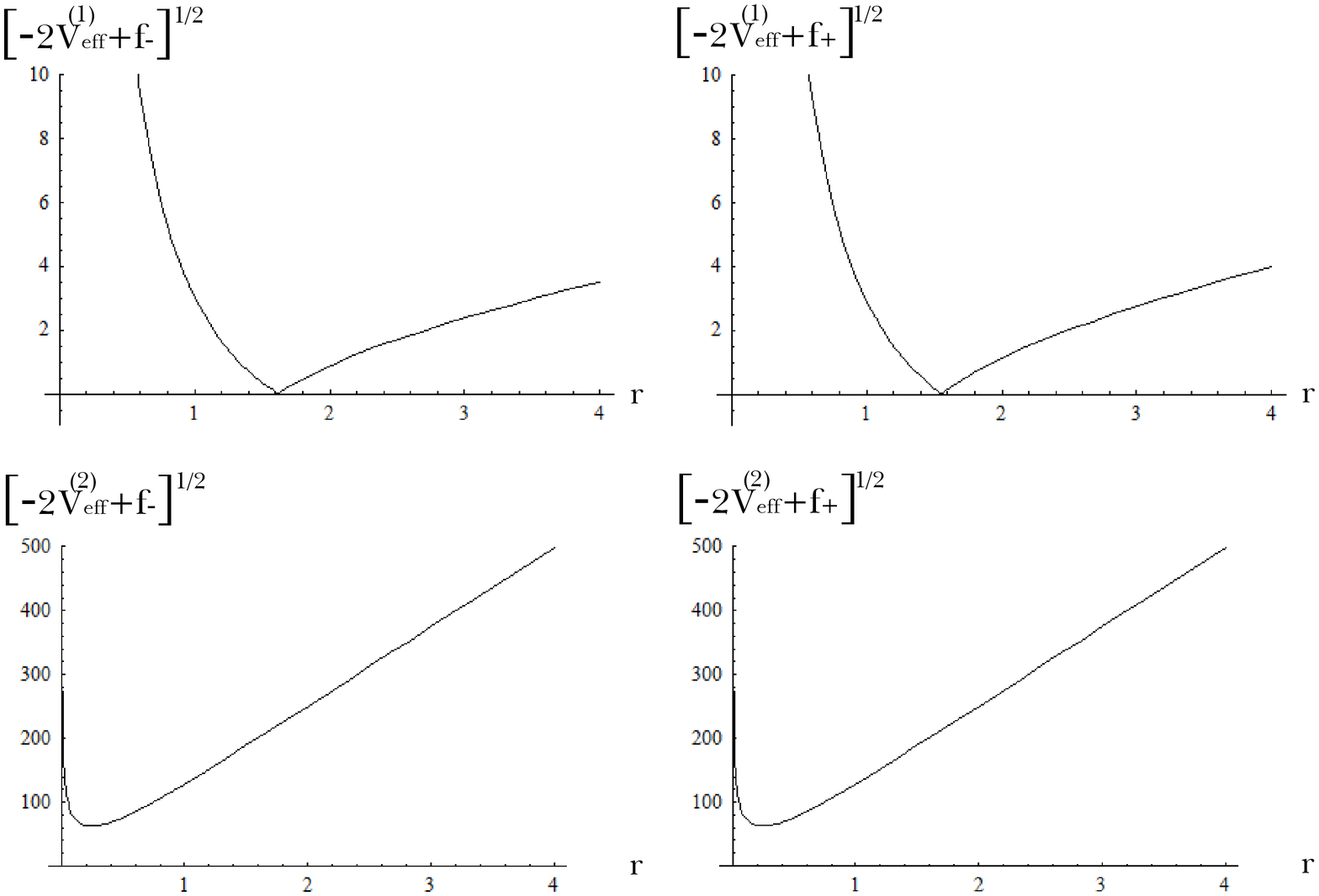}
\caption{\label{fig:less_sqrt_001}$\sqrt{\dot{r}^{2}+f_{-}}=\sqrt{-2V_{\mathrm{eff}}^{(1)}+f_{-}}$ (upper left), $\sqrt{-2V_{\mathrm{eff}}^{(1)}+f_{+}}$ (upper right), $\sqrt{-2V_{\mathrm{eff}}^{(2)}+f_{-}}$ (lower left), and $\sqrt{-2V_{\mathrm{eff}}^{(2)}+f_{+}}$ (lower right) for the $l=2$, $M=0.5$, $\sigma=0.01$, and $\Phi = 0.999$ case.}
\end{center}
\end{figure}

\begin{figure}
\begin{center}
\includegraphics[scale=0.45]{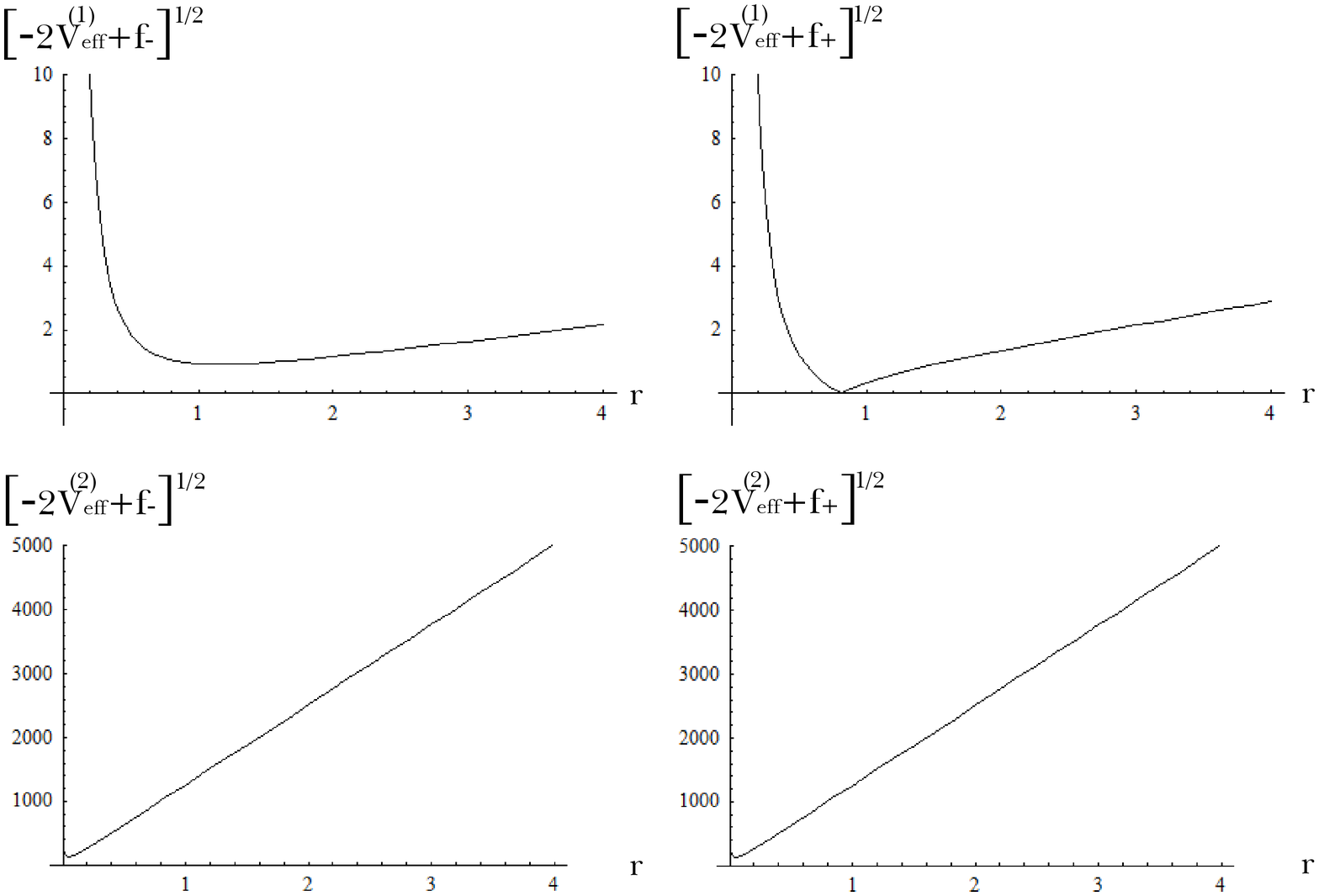}
\caption{\label{fig:less_sqrt_01}$\sqrt{-2V_{\mathrm{eff}}^{(1)}+f_{-}}$ (upper left), $\sqrt{-2V_{\mathrm{eff}}^{(1)}+f_{+}}$ (upper right), $\sqrt{-2V_{\mathrm{eff}}^{(2)}+f_{-}}$ (lower left), and $\sqrt{-2V_{\mathrm{eff}}^{(2)}+f_{+}}$ (lower right) for the $l=2$, $M=0.5$, $\sigma=0.1$, and $\Phi = 0.999$ case.}
\end{center}
\end{figure}

\begin{figure}
\begin{center}
\includegraphics[scale=0.4]{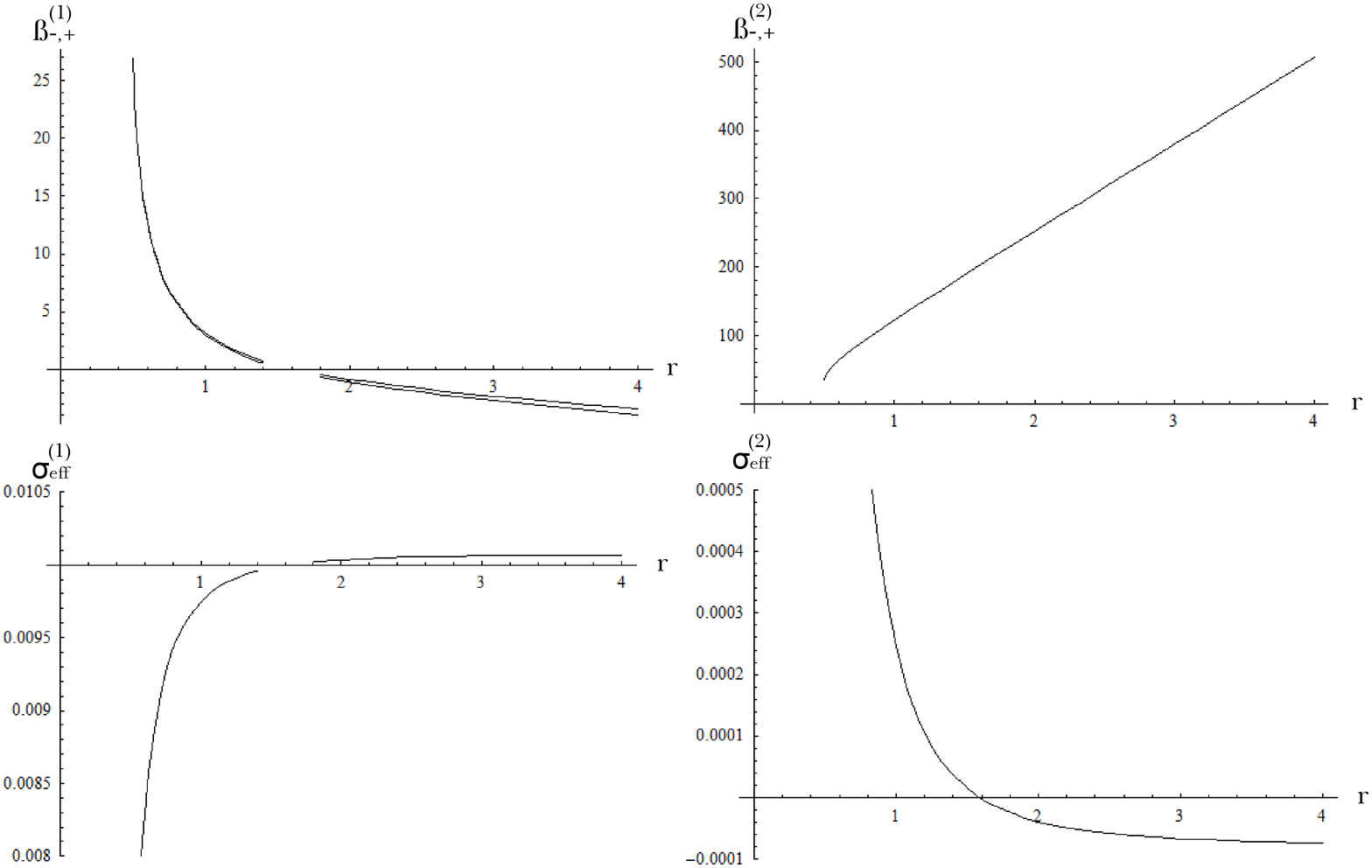}
\caption{\label{fig:more_tension_001}Effective extrinsic curvatures $\beta^{(1)}_{\pm}$ (upper left) change their sign from $+$ to $-$. Upper right is $\beta^{(2)}_{\pm}$, while two curves are folded. The effective tension $\sigma_{\mathrm{eff}}^{(1)}$ (lower left) is always positive. The effective tension $\sigma_{\mathrm{eff}}^{(2)}$ approaches a negative value. Here, $l=2$, $M=0.5$, $\sigma=0.01$, and $\Phi = 1.001$. Note that we do not plot $1.4 \lesssim r \lesssim 1.8$ for $V^{(1)}_{\mathrm{eff}}$ because the region is not allowed.}
\end{center}
\end{figure}

\begin{figure}
\begin{center}
\includegraphics[scale=0.4]{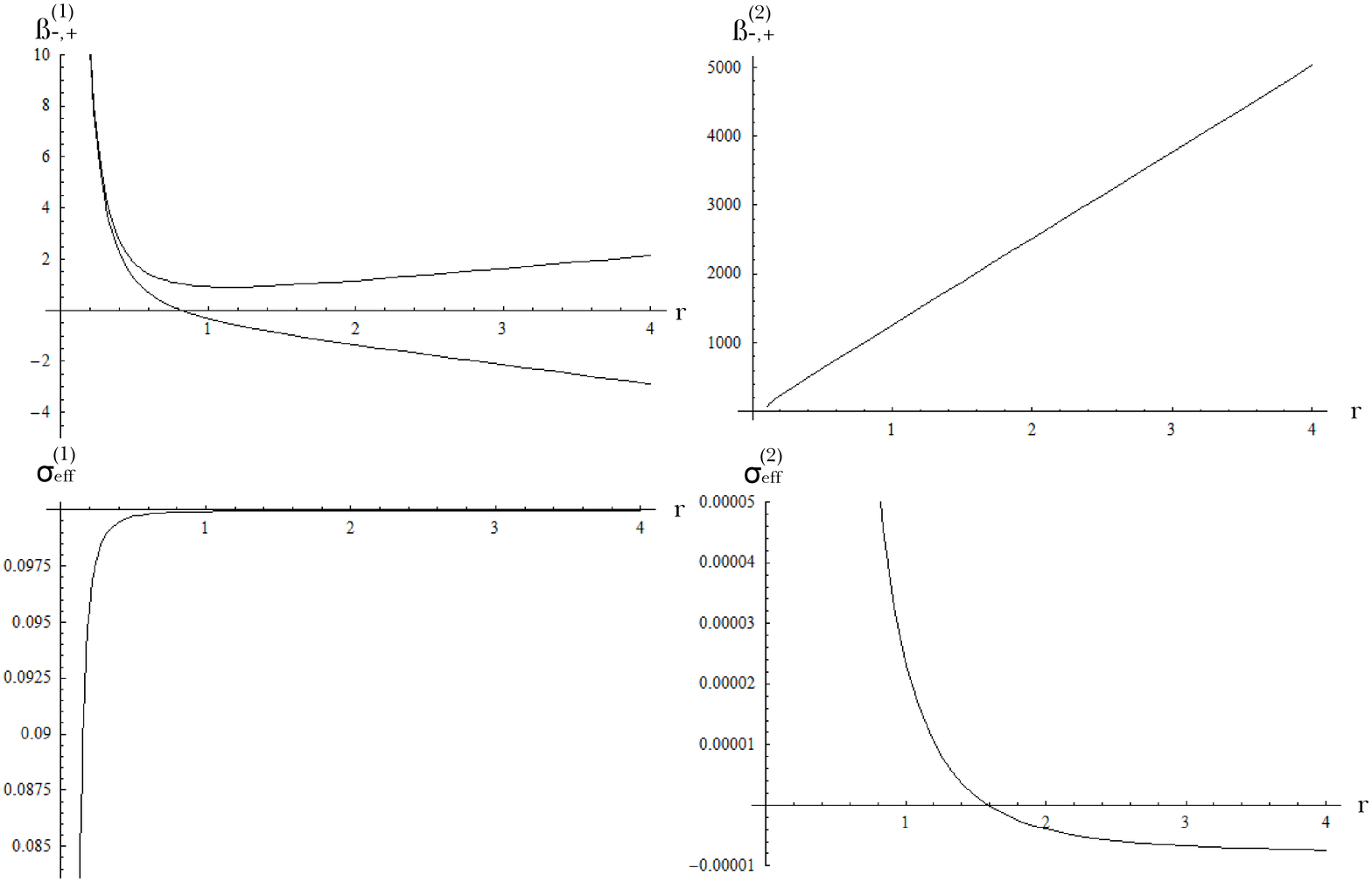}
\caption{\label{fig:more_tension_01}Effective extrinsic curvatures $\beta^{(1)}_{\pm}$ (upper left, the upper curve is $\beta^{(1)}_{-}$) and $\beta^{(2)}_{\pm}$ (upper right, two curves are folded). Effective tensions $\sigma_{\mathrm{eff}}^{(1)}$ (lower left) and $\sigma_{\mathrm{eff}}^{(2)}$ (right) are plotted. Here, $l=2$, $M=0.5$, $\sigma=0.1$, and $\Phi = 1.001$.}
\end{center}
\end{figure}

\begin{figure}
\begin{center}
\includegraphics[scale=0.4]{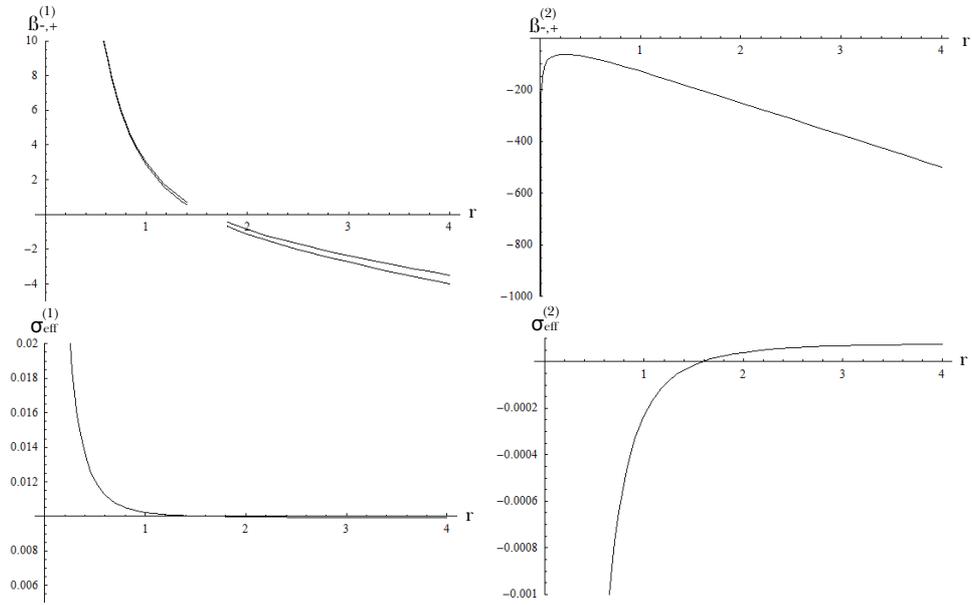}
\caption{\label{fig:less_tension_001}Effective extrinsic curvatures $\beta^{(1)}_{\pm}$ (upper left) change their sign from $+$ to $-$. Upper right is $\beta^{(2)}_{\pm}$, while two curves are folded. The effective tension $\sigma_{\mathrm{eff}}^{(1)}$ (lower left) is always positive. The effective tension $\sigma_{\mathrm{eff}}^{(2)}$ changes the value from a negative value to a positive value. Here, $l=2$, $M=0.5$, $\sigma=0.01$, and $\Phi = 0.999$. Note that we do not plot $1.4 \lesssim r \lesssim 1.8$ for $V^{(1)}_{\mathrm{eff}}$ because the region is not allowed.}
\end{center}
\end{figure}

\begin{figure}
\begin{center}
\includegraphics[scale=0.4]{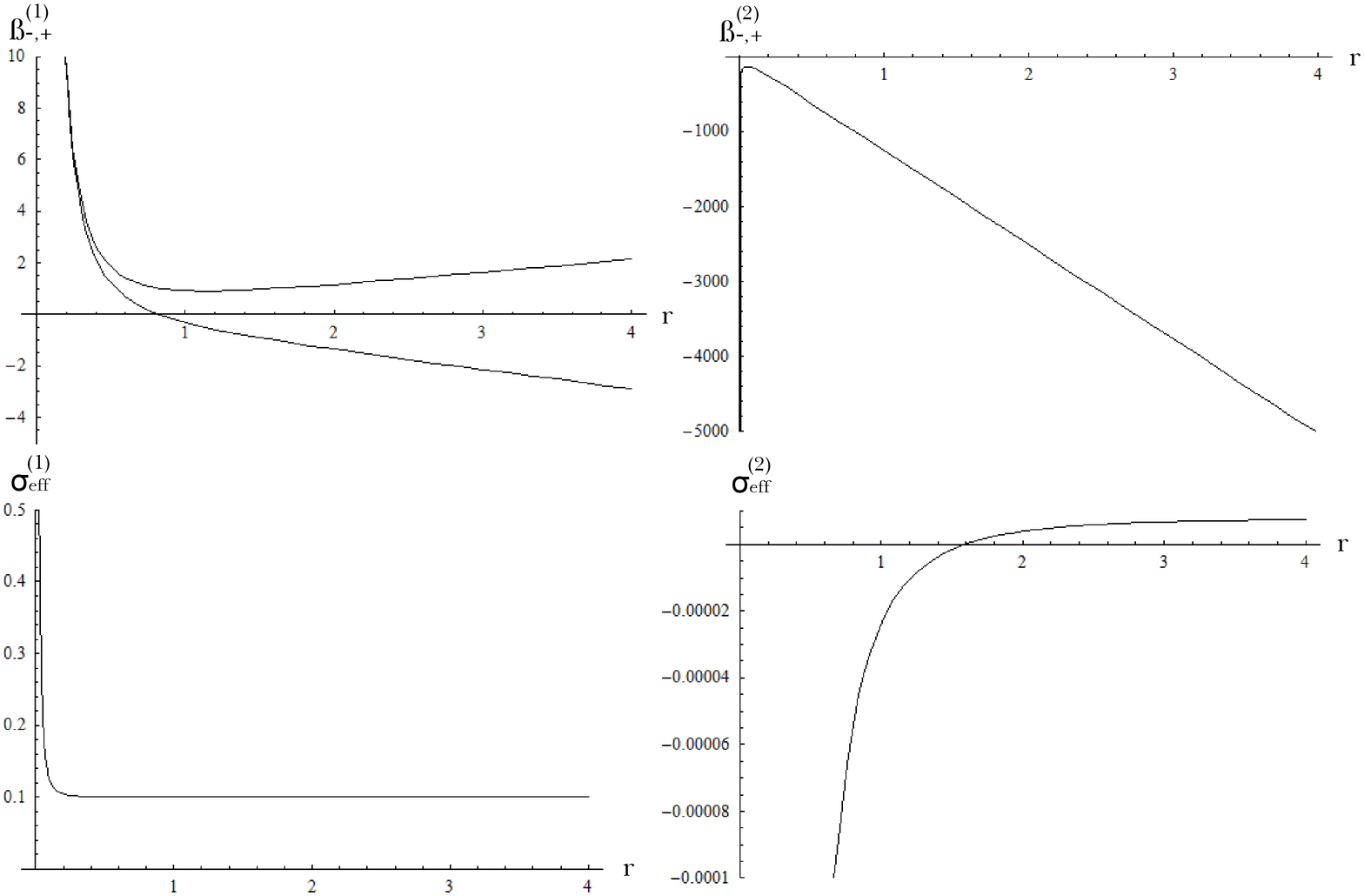}
\caption{\label{fig:less_tension_01}Effective extrinsic curvatures $\beta^{(1)}_{\pm}$ (upper left, the upper curve is $\beta^{(1)}_{-}$) and $\beta^{(2)}_{\pm}$ (upper right, two curves are folded). Effective tensions $\sigma_{\mathrm{eff}}^{(1)}$ (lower left) and $\sigma_{\mathrm{eff}}^{(2)}$ (right) are plotted. Here, $l=2$, $M=0.5$, $\sigma=0.1$, and $\Phi = 0.999$.}
\end{center}
\end{figure}

\begin{figure}
\begin{center}
\includegraphics[scale=0.4]{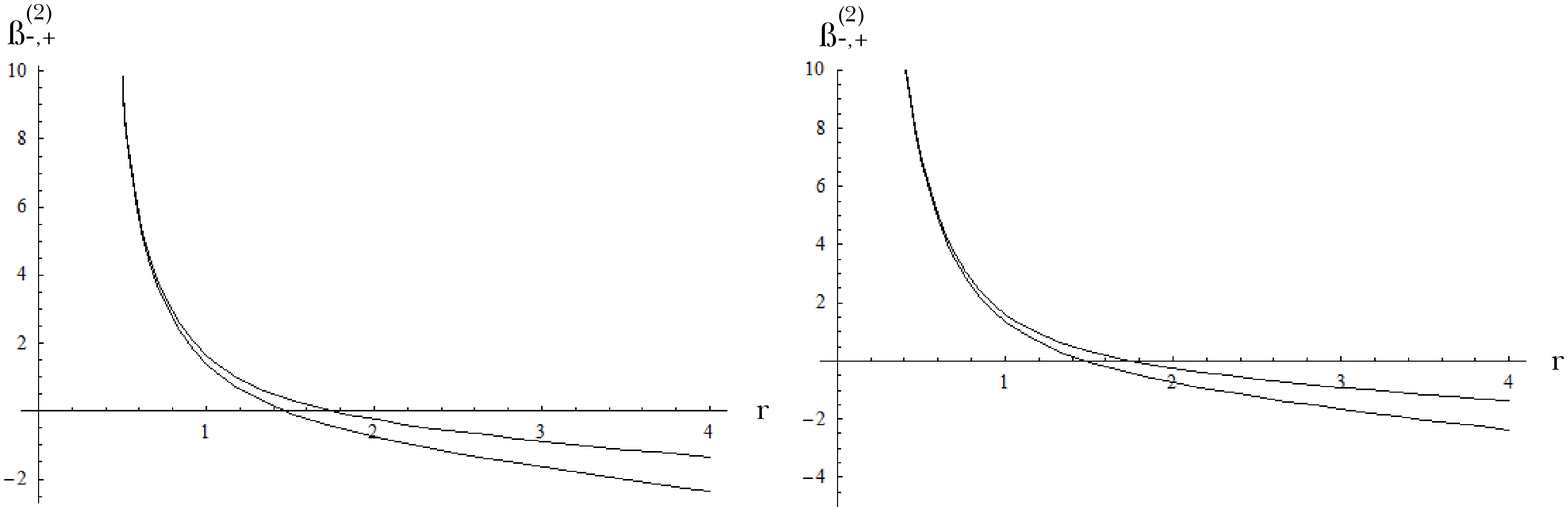}
\caption{\label{fig:wrong}Wrong choices of $\epsilon_{-}$ change the sign of $\beta^{(2)}_{\pm}$. Left is for $l=2$, $M=0.5$, $\sigma=0.01$, and $\Phi = 1.001$ and right is for $l=2$, $M=0.5$, $\sigma=0.01$, and $\Phi = 0.999$.}
\end{center}
\end{figure}

Now we briefly summarize how to determine the sign of $\epsilon_{-}$.
\begin{enumerate}
\item For the $V^{(1)}_{\mathrm{eff}}$ case, the sign $\epsilon_{\pm}$ will follow the Einstein limit.

Then, for $\sigma=0.01$, as mentioned in the previous paragraph, the symmetric collapsing solution is $\epsilon_{-}=1$ and the symmetric repulsing solution is $\epsilon_{-}=-1$. For $\sigma=0.1$, $\sqrt{\dot{r}^{2} + f_{-}}$ always has non-zero values, as shown in Figures~\ref{fig:more_sqrt_01} and \ref{fig:less_sqrt_01}, and then asymmetric solutions are $\epsilon_{-}=1$ since $\beta^{(1)}_{-}$ cannot change its sign.

If the effective tension $\sigma^{(1)}_{\mathrm{eff}}$ under previous choices is always positive, our conclusion is self-consistent. We can check from the effective tensions in Figures~\ref{fig:more_tension_001}, \ref{fig:more_tension_01}, \ref{fig:less_tension_001}, and~\ref{fig:less_tension_01}.

\item For the $V^{(2)}_{\mathrm{eff}}$ case, it is entirely unclear whether it will follow the Einstein limit or not. To figure out $\epsilon_{\pm}$, we must first observe whether $\sqrt{\dot{r}^{2} + f_{\pm}}$ become $0$ or not. If they do not become $0$, they will maintain their signs.

This can be confirmed by looking at the lower left and right of Figures~\ref{fig:more_sqrt_001}, \ref{fig:more_sqrt_01}, \ref{fig:less_sqrt_001}, and~\ref{fig:less_sqrt_01}.

\item Then, first choose $\epsilon_{-}$, and insert it into $\sigma^{(2)}_{\mathrm{eff}}$. Then, observe $\beta^{(2)}_{\pm}$; if they change their signs, the original choice of $\epsilon_{-}$ was inconsistent. Using this method, we can find a consistent $\epsilon_{-}$.

If we choose $\epsilon_{-}=+1$ for the $\Phi >1$ case and $\epsilon_{-}=-1$ for the $\Phi <1$ case, we can see that $\beta^{(2)}_{-}$ has consistent signs (Figures~\ref{fig:more_tension_001}, \ref{fig:more_tension_01}, \ref{fig:less_tension_001}, and~\ref{fig:less_tension_01}). However, if we choose opposite signs for each case, $\beta^{(2)}_{-}$ will change their signs and this situation is inconsistent (Figure~\ref{fig:wrong}).

\end{enumerate}





Note that, for $V_{\mathrm{eff}}^{(2)}$, some cases allow a negative tension, as can be seen in the lower right of Figure~\ref{fig:more_tension_001}, the lower right of Figure~\ref{fig:more_tension_01}, the lower right of Figure~\ref{fig:less_tension_001}, and the lower right of Figure~\ref{fig:less_tension_01}. These situations correspond to the violation of the null energy condition. Then, the violation of the null energy condition will give new causal structures that were disallowed in pure Einstein theory.

\begin{table}[b]
\begin{center}
\begin{tabular}{c|c|c|c|c}
\hline
& \multicolumn{2}{c}{\;\;\;\;\;\;$\Phi_{-} > 1$\;\;\;\;\;\;} & \multicolumn{2}{|c}{\;\;\;\;\;\;$\Phi_{-} < 1$\;\;\;\;\;\;}\\
\hline
& \;\;$r \sim 0$\;\; & \;$r \rightarrow \infty$\; & \;\;$r \sim 0$\;\; & \;$r \rightarrow \infty$\; \\
\hline \hline
$\beta^{(1)}_{+}$ & $+$ & $-$ & $+$ & $-$\\
\hline
$\beta^{(1)}_{-}$ & $+$ & $\mp$ & $+$ & $\mp$\\
\hline
$\beta^{(2)}_{+}$ & $+$ & $+$ & $-$ & $-$\\
\hline
$\beta^{(2)}_{-}$ & $+$ & $+$ & $-$ & $-$\\
\hline
\end{tabular}
\caption{\label{table:signs}Summary of the signs of our initial conditions. For $\mp$, the $-$ sign is for $\sigma=0.01$ and the $+$ sign is for $\sigma=0.1$.}
\end{center}
\end{table}

\subsection{Classification of causal structures}

If $\Phi_{-} > 1$, for small $r$, there is a radius bound inside of the Schwarzschild radius where the thin shell bubble is not allowed. One may interpret this by saying that if the radius of the shell becomes smaller than the region, the shell should collapse to a singularity, although the thin shell approximation is no more than a good approximation for the region and hence the inside bubble may become an unstable one.

Using the information on the signs of $\epsilon_{\pm}$, we finally classify the causal structures. The following are general rules to determine causal structures \cite{Blau:1986cw, Aguirre:2005xs, Freivogel:2005qh}:
\begin{itemize}
\item $r\sim 0$ limit
\begin{itemize}
\item If $\beta^{(i)}_{-}$ is positive, the outward-pointed normal has to point toward larger $r$; when the shell touches $r \sim 0$ in the de Sitter space, the shell has to touch the left boundary. If $\beta^{(i)}_{-}$ is negative, the shell has to touch the right boundary.
\item If $\beta^{(i)}_{+}$ is positive, when the shell touches $r \sim 0$ in the Schwarzschild space, the shell starting towards the right from the past singularity or moving towards the left before hitting the future singularity is allowed. If $\beta^{(i)}_{+}$ is negative, the opposite behavior will be obtained.
\end{itemize}
\item $r \rightarrow \infty$ limit
\begin{itemize}
\item If $\beta^{(i)}_{-}$ is negative, the shell starts from the past infinity towards the right and ends at the future infinity veering left. If $\beta^{(i)}_{-}$ is positive, the opposite behavior will be obtained.
\item If $\beta^{(i)}_{+}$ is negative, when the shell expands, the shell has to touch the left boundary. If $\beta^{(i)}_{+}$ is positive, when the shell expands, the shell has to touch the right boundary.
\end{itemize}
\end{itemize}

\begin{figure}
\begin{center}
\includegraphics[scale=0.5]{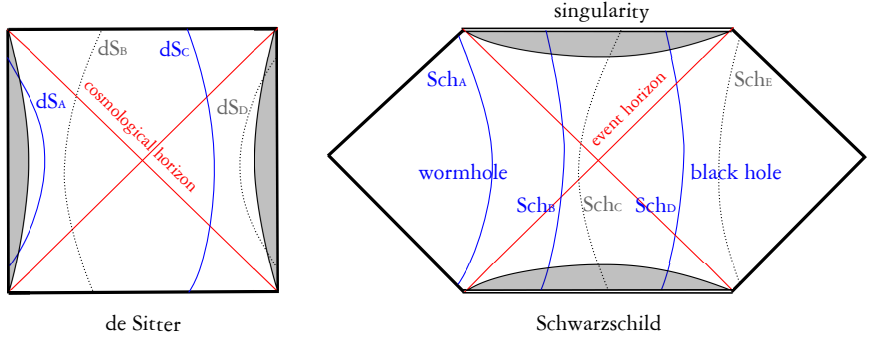}
\caption{\label{fig:thinshell}Solutions of the thin shell approximation (symmetric cases). Here, the black shadow region is the radius bound of the $\Phi_{-}>1$ case.}
\end{center}
\end{figure}

First, let us classify the symmetric solutions. The left diagram of Figure~\ref{fig:thinshell} is for the de Sitter space, and the right diagram is for the Schwarzschild space. For a collapsing case, $\mathrm{dS}_{\mathrm{A}}$ or $\mathrm{dS}_{\mathrm{D}}$ are possible; and $\mathrm{Sch}_{\mathrm{B}}$, $\mathrm{Sch}_{\mathrm{C}}$, or $\mathrm{Sch}_{\mathrm{D}}$ are possible. Also, for an expanding case, $\mathrm{dS}_{\mathrm{B}}$ or $\mathrm{dS}_{\mathrm{C}}$ are possible; and $\mathrm{Sch}_{\mathrm{A}}$ or $\mathrm{Sch}_{\mathrm{E}}$ are possible. However, according to the behavior of the extrinsic curvatures in $r\rightarrow 0$ or $r\rightarrow \infty$ limit, we can remove the solutions of $\mathrm{dS}_{\mathrm{D}}$, $\mathrm{Sch}_{\mathrm{C}}$, and $\mathrm{Sch}_{\mathrm{E}}$.
Also, $\mathrm{dS}_{\mathrm{B}}$ requires $\beta_{-} > 0$ in the $r \rightarrow \infty$ limit; however, it requires a sufficiently large tension, and then a symmetric solution will not be allowed.
Therefore, there are three possible solutions: $\mathrm{dS}_{\mathrm{A}}-\mathrm{Sch}_{\mathrm{B}}$, $\mathrm{dS}_{\mathrm{A}}-\mathrm{Sch}_{\mathrm{D}}$, and $\mathrm{dS}_{\mathrm{C}}-\mathrm{Sch}_{\mathrm{A}}$. However, these solutions are allowed in pure Einstein gravity.

\begin{figure}
\begin{center}
\includegraphics[scale=0.5]{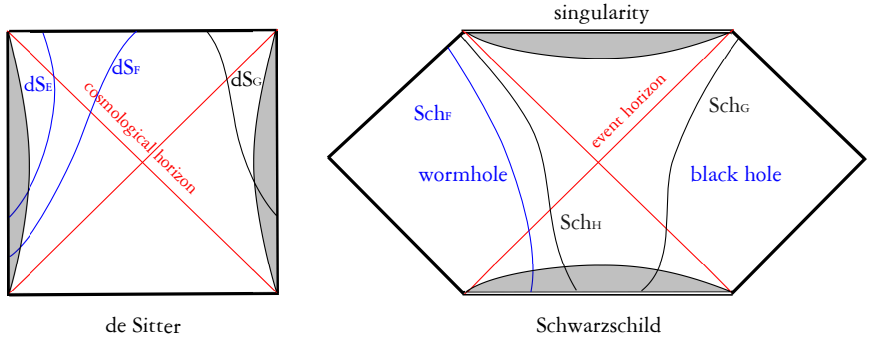}
\caption{\label{fig:thinshell2}Solutions of the thin shell approximation (asymmetric cases).}
\end{center}
\end{figure}

Second, let us classify the asymmetric solutions (Figure~\ref{fig:thinshell2}). The most interesting case is that of the creation of a bubble universe. In this case, we need to consider the example from an expanding to an expanding solution. Here, $\mathrm{dS}_{\mathrm{E}}$, $\mathrm{dS}_{\mathrm{F}}$ are allowed and, at that time, $\mathrm{Sch}_{\mathrm{F}}$ is allowed for positive tensions. $\mathrm{Sch}_{\mathrm{G}}$ is allowed from $V_{\mathrm{eff}}^{(2)}$ of the $\Phi_{-}>1$ case, but in this case, $\beta^{(2)}_{-}$ is always positive, and, hence, $\mathrm{dS}_{\mathrm{F}}$ corresponds. Also, $\mathrm{dS}_{\mathrm{G}}$ is allowed by $V_{\mathrm{eff}}^{(2)}$ of the $\Phi_{-}<1$ case with negative $\beta^{(2)}_{-}$ and, at that time, $\mathrm{Sch}_{\mathrm{H}}$ is allowed. Thus we have four possible solutions: $\mathrm{dS}_{\mathrm{E}}-\mathrm{Sch}_{\mathrm{F}}$, $\mathrm{dS}_{\mathrm{F}}-\mathrm{Sch}_{\mathrm{F}}$, $\mathrm{dS}_{\mathrm{F}}-\mathrm{Sch}_{\mathrm{G}}$, and $\mathrm{dS}_{\mathrm{G}}-\mathrm{Sch}_{\mathrm{H}}$.

\begin{figure}
\begin{center}
\includegraphics[scale=0.5]{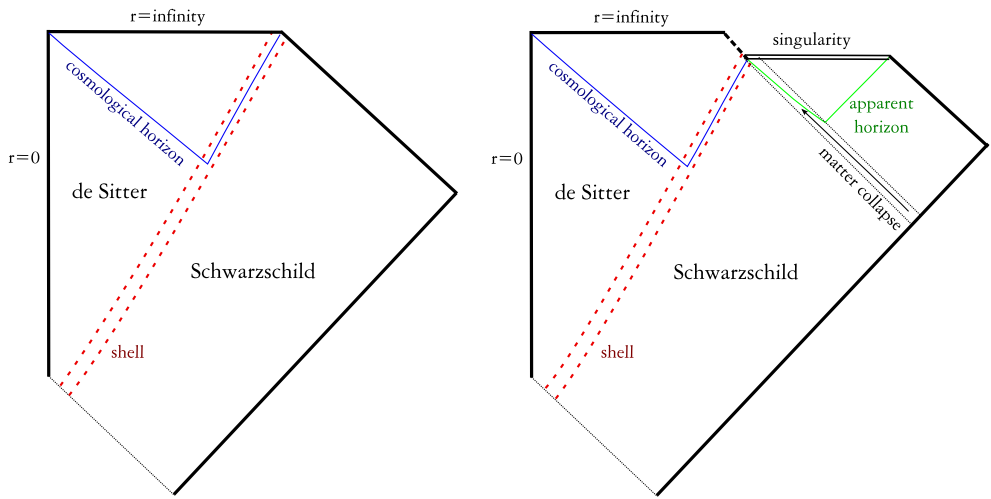}
\caption{\label{fig:conclusion}The causal structure of $\mathrm{dS}_{\mathrm{F}}-\mathrm{Sch}_{\mathrm{G}}$ (left). If we add sufficient energy to the shell, one can make a black hole and can separate the inside bubble universe from the outside (right).}
\end{center}
\end{figure}

Solutions $\mathrm{dS}_{\mathrm{E}}-\mathrm{Sch}_{\mathrm{G}}$ from $V_{\mathrm{eff}}^{(2)}$ of the $\Phi_{-}>1$ case as well as $\mathrm{dS}_{\mathrm{G}}-\mathrm{Sch}_{\mathrm{H}}$ from $V_{\mathrm{eff}}^{(2)}$ of the $\Phi_{-}<1$ case are new solutions that were disallowed in Einstein gravity. Especially, the former will be useful to discuss the information loss problem. If the bubble becomes larger than the cosmological horizon of the inside de Sitter space, it begins to inflate.
Then, there should be a neck of a wormhole on the shell. Although the thin shell approximation cannot describe the throat, it has been demonstrated by numerical simulations by previous authors \cite{Hansen:2009kn}. We note the causal structures in Figure~\ref{fig:conclusion}.


\subsection{A limiting case: $\Phi_{-}>1$ and $M=0$}

As a limiting case, we observe the $\Phi_{-}>1$ and $M=0$ case. We used $l=2$, $M=0$, $\sigma=0.01$, and $\Delta \Phi = 0.01$ (i.e., $\Phi_{-}=1.01$) so that
\begin{equation}
\epsilon \sim \frac{(\Delta \Phi)^{2}}{\sigma} \sim 10^{-2} \ll 1 \sim l.
\end{equation}
Hence, the thin shell approximation holds for these parameters.

\begin{figure}
\begin{center}
\includegraphics[scale=0.5]{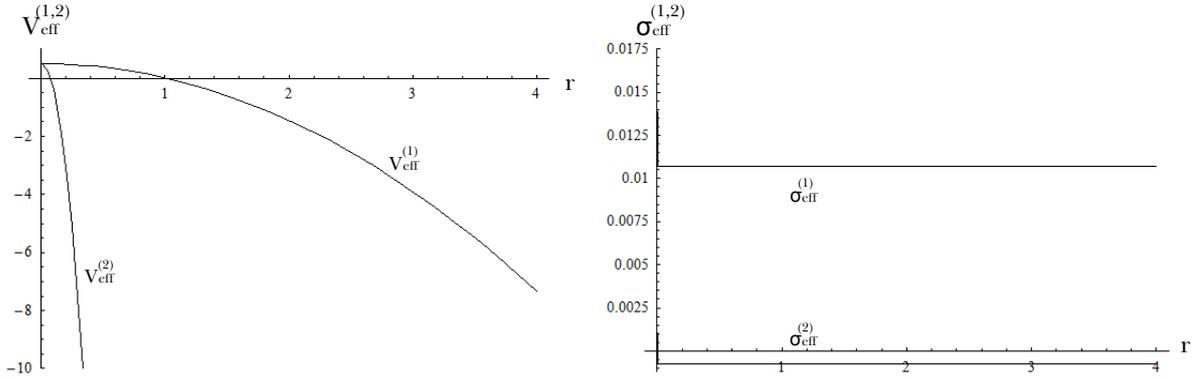}
\caption{\label{fig:more_M0}The effective potentials $V_{\mathrm{eff}}^{(1)}$ and $V_{\mathrm{eff}}^{(2)}$ (left) and effective tensions (right) for the $l=2$, $M=0$, $\sigma=0.01$, and $\Phi = 1.01$ case.}
\end{center}
\end{figure}
\begin{figure}
\begin{center}
\includegraphics[scale=0.5]{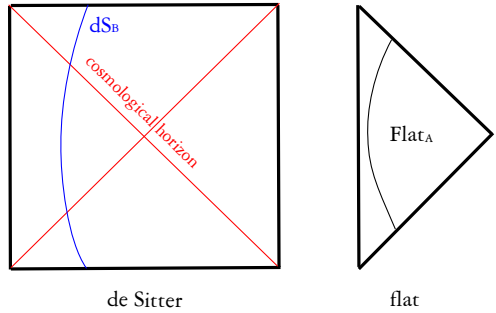}
\caption{\label{fig:thinshell3}Solutions of the thin shell approximation in a flat background. $\beta^{(2)}_{\mathrm{+}}$ is always positive in $r \rightarrow \infty$ limit. Therefore, $\mathrm{Flat}_{\mathrm{A}}$ is allowed for $V_{\mathrm{eff}}^{(2)}$.}
\end{center}
\end{figure}

We observed effective potentials $V_{\mathrm{eff}}^{(1)}$ and $V_{\mathrm{eff}}^{(2)}$, and the effective tensions $\sigma_{\mathrm{eff}}^{(1)}$ and $\sigma_{\mathrm{eff}}^{(2)}$ (Figure~\ref{fig:more_M0}).
All effective potentials monotonely decrease, so that symmetric bounce solutions are allowed.

We can think that the signs of $\epsilon_{-}$ will follow those of the results of previous sections, $\epsilon^{(1)}_{-}=-1$ and $\epsilon^{(2)}_{-}=+1$. Also, it is not difficult to check by using small and non-zero $M$ and using the previous procedures to determine the sign.

For the potential $V_{\mathrm{eff}}^{(1)}$, the effective tension $\sigma^{(1)}_{\mathrm{eff}}$ converges to a positive value. Then, in the flat background, it cannot touch the right boundary. This implies that such a bubble cannot be described by the thin shell approximation, and that the bubble is unstable \cite{Hansen:2009kn}. For the potential $V_{\mathrm{eff}}^{(2)}$, the effective tension $\sigma^{(2)}_{\mathrm{eff}}$ converges to a negative value. Then, in the flat background, it can touch the right boundary. Then, we obtain one interesting solution (Figure~\ref{fig:thinshell3}). We obtain $\mathrm{dS}_{\mathrm{B}}-\mathrm{Flat}_{\mathrm{A}}$ since $\beta^{(2)}_{-}$ is always positive; this was disallowed in pure Einstein gravity with the null energy condition.

Note that if we choose $\Phi_{-}<1$, again, only symmetric expanding solutions will be allowed. However, in this case, $\beta_{+}$ should be negative, and, hence, stable solutions are not allowed.

\section{\label{sec:dis}Discussion}

In this paper, we study the dynamics of false vacuum bubbles in the Brans-Dicke theory using the thin shell approximation.
We considered a false vacuum bubble that has different values of the Brans-Dicke field between the inside false vacuum region and the outside true vacuum region.
We observed that the Brans-Dicke theory allows two effective potentials; one is similar to that of the Einstein case while the other has non-trivial properties. Detailed classifications of the extrinsic curvatures and the causal structures are new to this analysis and were not fully discussed in the previous work about a non-minimally coupled case \cite{lll2006, lll2008, lll20082}.

In this paper, the given tension $\sigma$ is a function of $\omega$. If $\omega$ is sufficiently large, we may lose the condition for the thin shell approximation. (Further studies on the shell where the Brans-Dicke field smoothly transits from inside to outside will be discussed in the future paper of the authors \cite{kllly}.) For observational tests, it is known that the value of $\omega$ should be greater than $4 \times 10^{4}$ \cite{Ber}. However, in various physical situations, small $\omega$ parameters can be allowed. Even though the small $\omega$ is not for our universe, small $\omega$ can be allowed by the fundamental theory \cite{Gasperini:2007zz, Randall:1999ee, Garriga:1999yh, Fujii:2003pa}, and such small value of $\omega$ may have some implications, e.g., a violation of energy conditions or a violation of unitarity which were impossible in Einstein gravity. Moreover, even in the presence of small $\omega$, it may be possible to find a viable model for our observational tests \cite{Khoury:2003aq}.

Within a certain limit of field values, the difference of field values makes the effective tension of the shell negative and induces a violation of the null energy condition. This allows new expanding solutions, which were disallowed in Einstein gravity, that reach the outside of a Schwarzschild wormhole (Figure~\ref{fig:thinshell}, \ref{fig:thinshell2}, and \ref{fig:thinshell3}) \cite{Blau:1986cw, Aguirre:2005xs, Freivogel:2005qh}, especially, $\mathrm{dS}_{\mathrm{F}}-\mathrm{Sch}_{\mathrm{G}}$ for potential $V_{\mathrm{eff}}^{(2)}$ with $\Phi_{-}>1$.
The physical meaning is that a small false vacuum bubble expands forever but the bubble is outside of a Schwarzschild black hole.
If the bubble becomes larger than the cosmological horizon of the inside de Sitter space, it begins to inflate and there should be a future infinity that is separated from the outside space (Figure~\ref{fig:conclusion}).

If these bubbles are able to be prepared by physical processes, it can cause a violation of unitarity or a loss of information. One question is whether the initial states can be prepared or not. If we assume Einstein gravity, global hyperbolicity, and the null energy condition, an initial state of an inflating space cannot be geodesically complete along the past direction \cite{Farhi:1986ty}. Hence, the initial condition requires more than general relativity \cite{Aguirre:2005xs, Farhi:1989yr, Freivogel:2005qh}. However, for our new solutions, the previous argument cannot have this implication, since the bubbles violate the null energy condition. Therefore, there is no reason to believe that the bubbles are unbuildable.

Thus, if one accepts that our new expanding bubble solutions are allowed so that the solutions can violate unitarity and cause an information loss problem, it will have important implications \cite{Yeom:2009mn}. For example, even if the background is an anti de Sitter space or a de Sitter space, as long as the background cosmological constant is sufficiently small, our results will be maintained. Then, one may infer that the Brans-Dicke theory in an anti de Sitter or a de Sitter space may violate unitarity. What is the implication of this situation to the AdS/CFT or dS/CFT correspondence \cite{Maldacena:1997re}? What is the implication to the information loss problem of black holes \cite{inforpara}? These problems remain for future work.

\section*{Acknowledgments}

DY would like to thank Ewan Stewart for discussions and encouragement.
WL would like to thank Hongsu Kim for helpful discussions.
DY was supported by Korea Research Foundation grants (KRF-313-2007-C00164, KRF-341-2007-C00010) funded by the Korean government (MOEHRD) and BK21.
BHL and WL were supported by a Korea Science and Engineering Foundation (KOSEF) grant funded by the Korean government (MEST) through the Center for Quantum Spacetime (CQUeST) of Sogang University with grant number R11 - 2005 - 021.
%WL was supported by a Korea Research Foundation Grant funded by the Korean Government (MOEHRD) (KRF-2007-355-C00014).
WL was supported by the National Research Foundation of Korea Grant funded by the Korean Government (Ministry of Education, Science and Technology)[NRF-2010-355-C00017].

\end{document}